\documentclass[%
 preprint,
 amsmath,amssymb,
 aps,
]{revtex4-1}

\usepackage{graphicx}
\usepackage{bm}

\usepackage{amsthm,amsfonts,amssymb,epsfig,graphics,amsmath,oldgerm,color}

\usepackage[latin1]{inputenc}

\newcommand{\R}{{\mathbb R}}
\newcommand{\N}{{\mathbb N}}
\newcommand{\Z}{{\mathbb Z}}

\newcommand{\eps}{\varepsilon}

\newcommand{\pa}{\partial}
\newcommand{\bspm}{\left(\begin{smallmatrix}}\newcommand{\espm}{\end{smallmatrix}\right)}
\newcommand{\bpm}{\left(\begin{matrix}}\newcommand{\epm}{\end{matrix}\right)}

\newcommand{\PROOF}{\textbf{Proof.} }

\def\epsilon{\varepsilon}
    
\def\beq{\begin{equation}}
\def\eeq{\end{equation}}

\begin{document}


\title{Dark Breathers in Granular Crystals}


\author{C. Chong}
\homepage{http://www.math.umass.edu/~chong}
\email{chong@math.umass.edu}
 \affiliation{Department of Mathematics and Statistics, University of Massachusetts \\
Amherst MA 01003-9305, USA}  

\author{P.G. Kevrekidis}
\affiliation{Department of Mathematics and Statistics, University of Massachusetts \\
Amherst MA 01003-9305, USA}%

\author{G. Theocharis}%
 \affiliation{Laboratoire {d'Acoustique} de {l'Universit\'e} du Maine, UMR-CNRS 6613 \\  Av. Olivier Messiaen, 72085 Le Mans, France}%

\author{Chiara Daraio}
\affiliation{Engineering and Applied Sciences, California Institute of Technology \\ Pasadena, CA 91125, USA}

\date{\today}

\begin{abstract}
We present a study of the existence, stability and bifurcation structure of 
families of {\it dark} breathers in a one-dimensional uniform chain of spherical beads under static load.  A defocusing nonlinear Schr\"odinger equation (NLS) is derived for frequencies that are close to the edge of the phonon band and is used to construct targeted initial conditions for numerical computations. 
Salient features of the system include the existence of large amplitude solutions that bifurcate with 
the small amplitude solutions described by the NLS equation, and 
the presence of a nonlinear instability that, to the best of the authors knowledge, has not been observed in
classical Fermi-Pasta-Ulam lattices.
 Finally,  it is also demonstrated that these dark breathers can be detected in a
 physically realistic way by merely actuating the ends of an 
initially at rest chain of beads and inducing destructive interference between their signals. 
\begin{description}
\item[PACS numbers]
63.20.Ry, 63.20.Pw, 45.70.-n, 46.40.-f

\end{description}
\end{abstract}

\pacs{Valid PACS appear here}
\maketitle



\section{Introduction}\label{intro}

Granular crystals, which consist of closely packed arrays of elastically interacting particles, 
have sparked recent theoretical and experimental interest. On one hand, the relevant mathematical
model is difficult to analyze due to, among other things, the lack of smoothness of the corresponding
potential function, calling for new ideas to study this nonlinear lattice 
problem~\cite{pego,atanas}. On the other hand,
the ability to conduct experiments for a wide range of setups, including chains consisting of beads of various geometries and masses, nonlinearity strength,
and even dimension \cite{nesterenko1,sen08,nesterenko2,coste97}, make the subject of granular crystals exciting for its potential relevance in applications
including shock and energy absorbing 
layers~\cite{dar06,hong05,fernando,doney06}, 
actuating devices \cite{dev08}, acoustic lenses \cite{Spadoni}, acoustic diodes \cite{china,Nature11} and sound scramblers \cite{dar05,dar05b}, to name a few. 

\begin{figure}
 \centerline{\epsfig{file=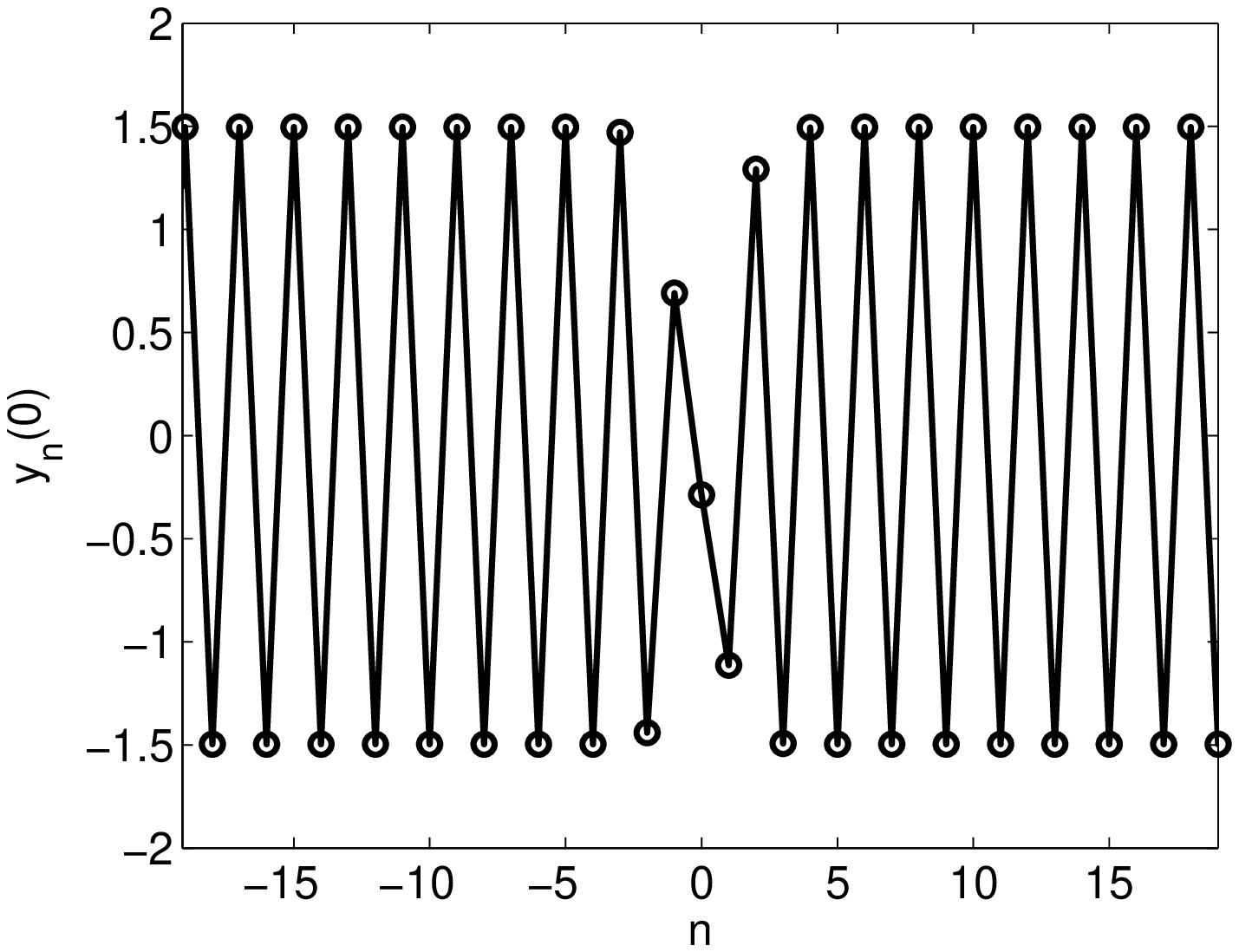,width=.3 \textwidth}
\epsfig{file=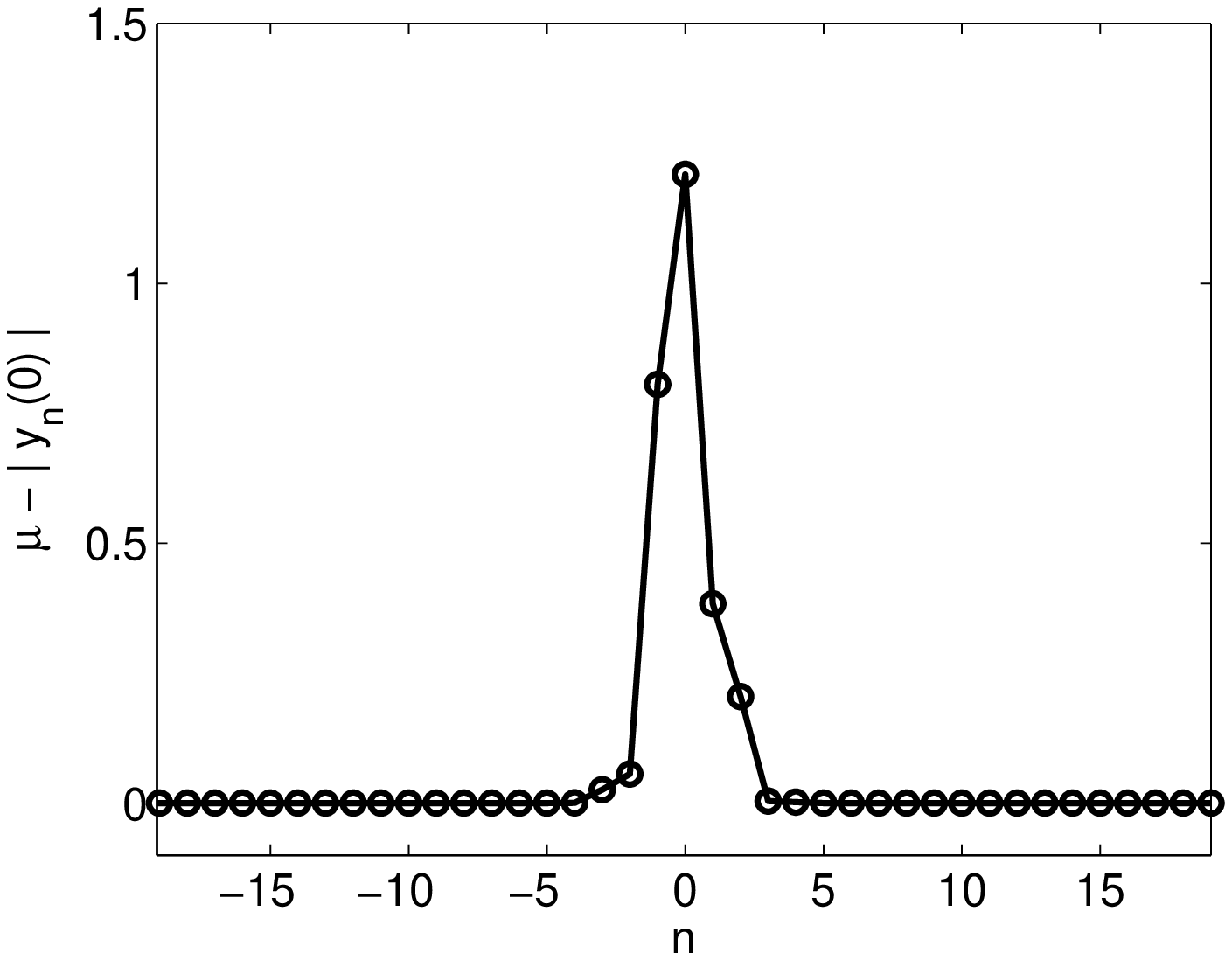,width=.3 \textwidth}
\epsfig{file=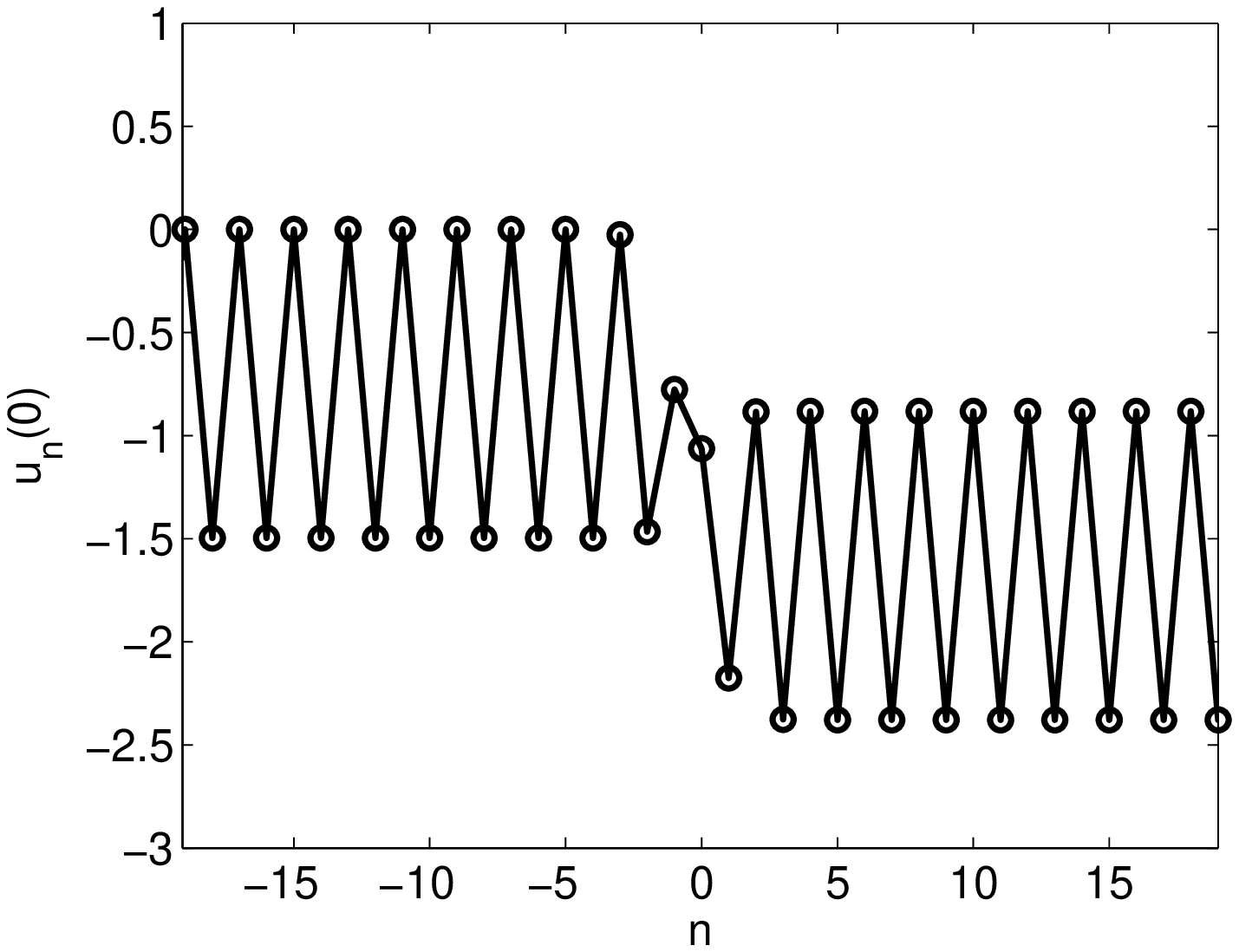,width=.3 \textwidth}
 }
 \centerline{\epsfig{file=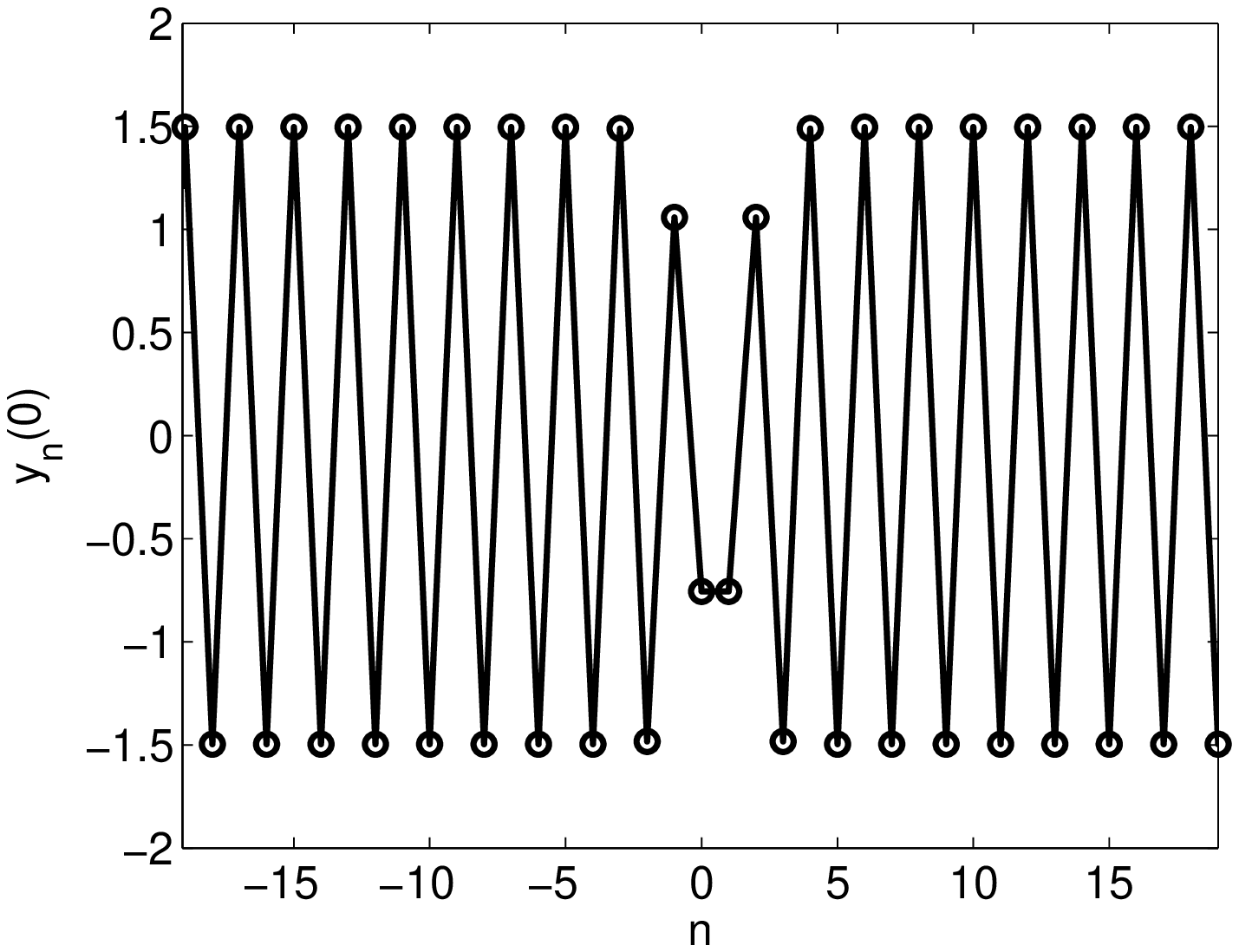,width=.3 \textwidth}
\epsfig{file=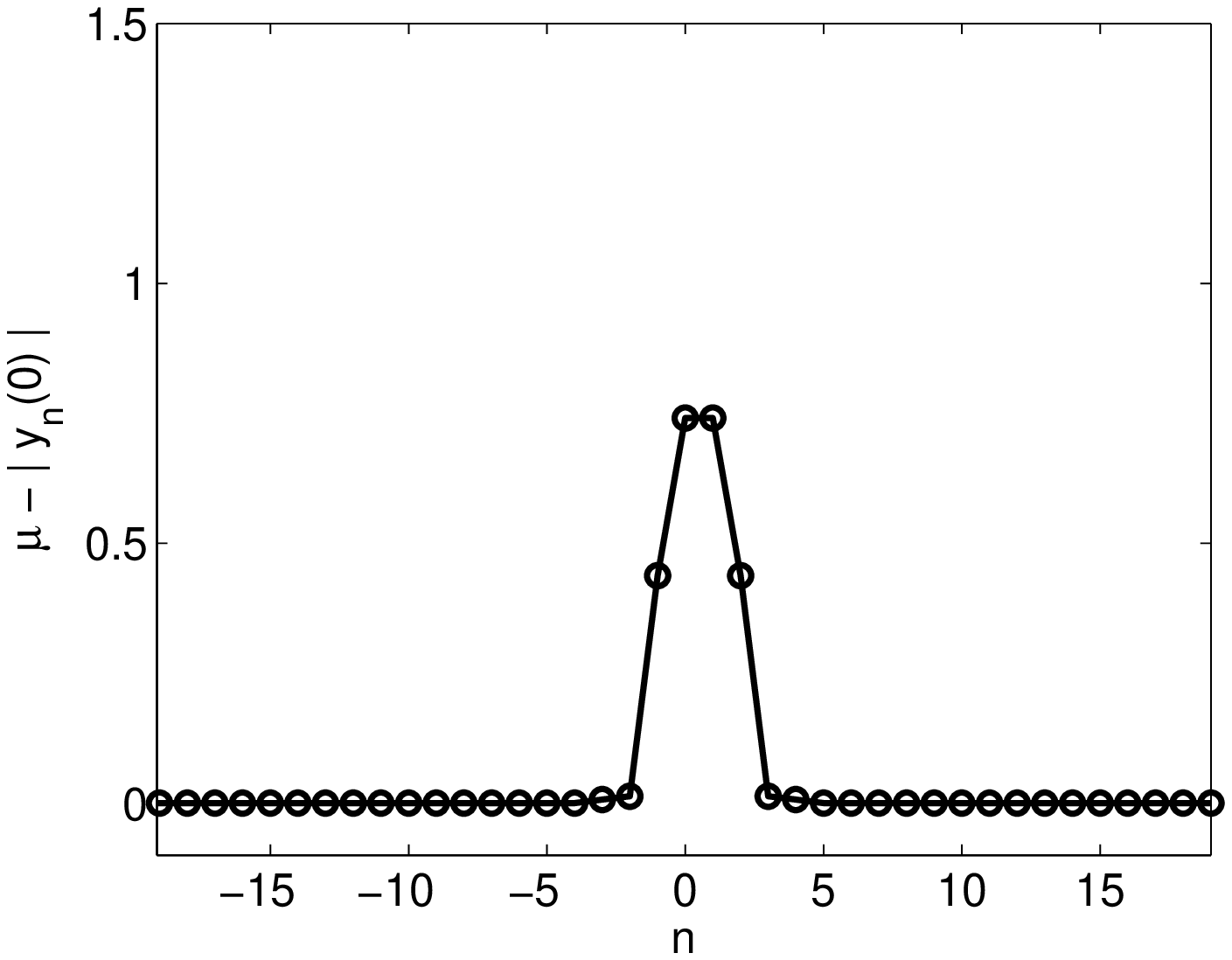,width=.3 \textwidth}
\epsfig{file=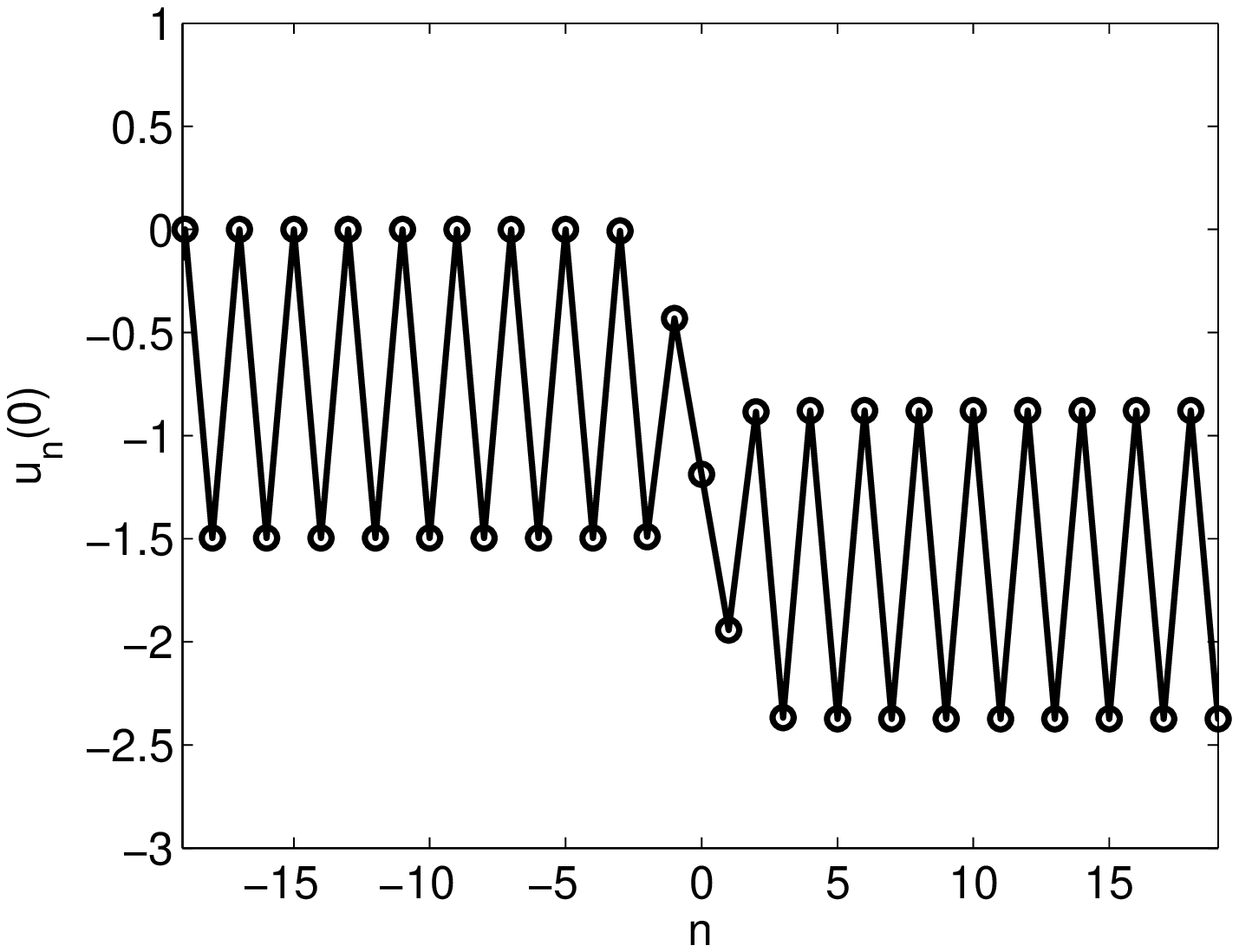,width=.3 \textwidth}
 }
\caption{Top: A  site-centered dark breather solution with $\omega_b = 1.9$ in the strain variable $y_n = u_{n+1} - u_n$ (left),
renormalized strain variable $\mu - | y_n |$, where $\mu$ is the breather amplitude (middle) and in displacement variables $u_n$ (right).
The markers indicate where the solution takes values on the lattice. The solid line is shown for clarity. Bottom:
Same as top row but for the bond-centered solution.}
\label{profiles}
\end{figure}

One fundamental structure known to exist in granular crystals is the so-called intrinsic localized mode, or discrete breather.
Discrete breathers are time-periodic solutions of the underlying equations of motion which are localized in space. Generally speaking, they
 are well studied, and are known to exist in a host of nonlinear lattice
models \cite{Flach2007}.  The most commonly studied breather is one with 
tails decaying to zero, which is often referred to as a bright breather. The 
term ``bright'' is used, as the solution can be viewed as a discrete carrier 
wave with amplitudes modulated by a bright soliton. It is then
natural to consider a discrete wave modulated by a dark soliton, which is 
then in turn called a dark breather, 
see Fig.~\ref{profiles} for an example. Although bright discrete breathers are known to exist in granular crystals
(in dimer or higher periodic configurations, and in monomer chains with defects for example \cite{Theo2009,Theo2010,Boechler2009,hooge11}), 
the properties of dark discrete breathers remains an open question, and is the subject of this work. 

It should be highlighted here that in monomer i.e.,
homogeneous chains
(for which, there is purely an acoustic band), the conditions established
for the bifurcation of discrete breathers~\cite{Flach2007} do not allow
the bifurcation of bright breathers. For that reason the latter have
only been identified in heterogeneous configurations with different
periodicities~\cite{Theo2010,Boechler2009,hooge11}, or in monomer
settings bearing defects~\cite{Theo2009}, where the localization is
not intrinsic to the chain but rather is supported as an impurity mode
by the defect. The {\it only} possibility that exists in the homogeneous
chain, under static load (i.e., in the linearizable limit) 
for intrinsic localization may arise in the form of the dark breathers
proposed herein. For this reason, we believe that such states are
fundamental ones in the study of granular systems and merit an
investigation of their existence properties, linear and nonlinear 
stability, as well as of their dynamics. These topics form the focus
of the present work which is structured as follows. In section 2, we
present the general properties of the model. In section 3, we analyze
the model near the band edge of its linear acoustic spectrum, and derive 
a defocusing nonlinear Schr{\"o}dinger equation characterizing its envelope
dynamics. In section 4, we specify the finite dimensional lattice dynamical
system to be studied and its bifurcation analysis is carried out in section 5.
While sections 6 and 8 consider variants of the center position of the breather
and of the associated boundary conditions, section 7 focuses on the linear
stability properties of the states. Section 9 analyzes the validity of
the analytical approximation, while section 10 makes connections with 
current experimental settings, presenting a
scheme for the potential realization of the dark breathers. 
Finally, section 11 presents our 
conclusions and some challenges for future work.

\section{Model}

The model describing the dynamics of a one dimensional (1D) homogeneous chain of spherical beads is given by \cite{nesterenko1,coste97}: 	
\begin{equation} \label{GC}
 M \ddot{u}_n =   V_{\mathrm{GC}}'( u_n - u_{n-1}) - V_{\mathrm{GC}}'( u_{n+1} - u_{n} ),
\end{equation}
with,
\begin{equation}
V_{\mathrm{GC}}'(x) =   A \left[ \delta_0 - x \right]_+^{3/2}, 
\label{chain}
\end{equation}
where $n \in I$, with $I$ a countable index set, 
$u_n=u_n(t)\in\R$ is the displacement of the $n$-th bead from equilibrium position at time $t$, $A$ is a material parameter (depending on the elastic
properties of the material and the geometric characteristics of the 
beads~\cite{nesterenko1}), $M$ is the bead mass and $\delta_0$ is an
equilibrium displacement induced by a static load $F_0=A 
\delta_0^{3/2}$. The bracket is defined by $[x]_+ = \mathrm{max}(0,x)$.
Equation~\eqref{GC} with $I=\Z$ has the Hamiltonian,
$$H = \sum_{n\in\N} \frac{1}{2} M v_n^2 + V_{\mathrm{GC}}(u_{n+1} - u_n), $$
where $v_n = \dot{u}_n$. Considering small amplitude
solutions, i.e., 
$$ \frac{|u_{n-1} - u_n|}{\delta_0} \ll 1,$$
implies that we can Taylor expand $V_{\mathrm{GC}}'(x)$ . Keeping terms up to fourth order yields an
approximate model,
\begin{equation}\label{FPU}
  M \ddot{u}_n = V_{\mathrm{FPU}}'(u_{n+1} - u_n) - V_{\mathrm{FPU}}'(u_n - u_{n-1} ),
\end{equation}
where 
$$ V_{\mathrm{FPU}}'(x) = K_2 x + K_3 x^2 + K_4 x^3,$$
and
\begin{equation} \label{GCcoeff}
 K_2 = \frac{3}{2}A\delta_0^{1/2}, \quad K_3 = -A\frac{3}{8} 
\delta_0^{-1/2},\quad K_4 = -A\frac{3}{48}\delta_0^{-3/2}.
\end{equation}
Equation~\eqref{FPU} is an example of the well studied Fermi-Pasta-Ulam (FPU) model. 
It is has been shown by James \cite{James01,James03} (see also
the discussion in~\cite{Flach2007}) 
that small amplitude bright breathers exist in the FPU model for frequencies above the phonon band if and only if $B := 3 K_2 K_4 - 4K_3^2 >0$. 
For coefficients such that $B<0$, small amplitude dark breathers were shown to exist for frequencies 
within the phonon band. The coefficients~\eqref{GCcoeff} associated to the granular crystal model
correspond to $B<0$, and thus the existence results for dark breathers can be carried
over for sufficiently small amplitudes (see e.g. Theorem 5 of \cite{James03}). 

In this paper, we begin with small amplitude solutions, where the phenomenology of the FPU lattice
and granular crystal lattice are very similar. However, we will see that as we depart from the familiar small 
amplitude limit, a host of new phenomena will emerge that are particular 
to the monomer granular crystal model and its rather special nonlinearity
bearing the unusual $3/2$ power, but also the tensionless characteristic
(namely, that there is only force between the beads when they are in contact).

\section{Derivation of the defocusing NLS equation in the strain variable formulation}
\label{derive}

It is more natural to derive the nonlinear Schr\"odinger equation in terms of the strain variable $y_n = u_{n+1} - u_n$.
We note that for a finite chain, a displacement variable solution can be recovered using the relation,
\begin{equation} \label{strain2dis}
 u_{n} = u_{1} + \sum_{j=1}^{n-1} y_j , 
\end{equation}
where an arbitrary choice for the first node must be made. Equation~\eqref{GC} has the following form in
the strain variables,
\begin{equation}\label{GC_strain}
 M\ddot{y}_n = - V_{\mathrm{GC}}'(y_{n+1}) + 2 V_{\mathrm{GC}}'(y_{n}) - V_{\mathrm{GC}}'(y_{n-1}). 
\end{equation}
The FPU equation can again be derived, yielding,
\begin{equation}\label{FPU_strain}
 M\ddot{y}_n = V_{\mathrm{FPU}}'(y_{n+1}) - 2 V_{\mathrm{FPU}}'(y_{n}) + V_{\mathrm{FPU}}'(y_{n-1}). 
\end{equation}
The linear problem (i.e. when $K_3 = K_4=0$) is solved by,
$$y_n(t) = y_{no} e^{i(k n + \omega t)},$$ 
for all $k\in \R$ where $\omega$ and $k$ are related through the dispersion relation,
$$\omega(k)^2 = 4 K_2/M\sin^{2}(k/2), $$
such that the cutoff point of the phonon band is $2 \sqrt{K_2/M}$.

Consider the standard NLS multiple scale ansatz, which up to first order has the form,
\begin{equation} \label{ansatz_strain}
y_n(t) \approx \psi_n(t) := \eps A(X,T) e^{i( k_0 n + \omega_0 t)}  + \mathrm{c.c.}  \,,  \quad X=\eps( n + c t), \quad T = \eps^2 t,
\end{equation}
where $\eps \ll 1$ is a small parameter, effectively parameterizing the
solution amplitude (and also its inverse width). The 
substitution of this ansatz into~\eqref{FPU_strain}
and equation of the various orders of $\epsilon$ leads to the dispersion relation 
$\omega_0 = \omega(k_0)$, 
the group velocity relation $c = \omega'(k_0)$, 
and the nonlinear Schr\"odinger equation,
\begin{equation}\label{NLS}
 i \pa_T A(X,T) + \nu_2 \pa_X^2 A(X,T) + \nu_3 A(X,T)|A(X,T)|^2 = 0,
\end{equation}
%
where $\nu_2 = -\omega''(k_0)/2 > 0$ and 
\begin{equation} \label{gamma}
\nu_3 = \frac{K_3^2}{K_2^2}\tilde{\gamma} + \frac{3 K_4}{2 K_2} \omega(k_0),
\end{equation}
while
\begin{eqnarray} \label{gamma1}
\tilde{\gamma} & = & \frac{\omega(k_0)}{2} \left(\frac{\omega(2 k_0)}{2 \omega(k_0)-\omega(2k_0) }
-\frac{\omega(2 k_0)}{2 \omega(k_0)+\omega(2k_0) } \right.
\\ && \left. \qquad \qquad +
\frac{2\omega'(0)}{\omega'(k_0)-\omega'(0) }
-\frac{2\omega'(0)}{\omega'(k_0)+\omega'(0) }
 \right). \nonumber
\end{eqnarray}
Full details of the derivation of the NLS equation starting from the homogeneous FPU equation, including the higher order terms of the ansatz,
can be found e.g. in \cite{Schn10}. Since we seek standing wave solutions, we choose the wavenumber to be at the edge of the phonon band $k_0 = \pi$, such that the group velocity vanishes, 
$\omega_0 = 2 \sqrt{K_2/M}$, and
$$\left.\nu_3\right|_{k_0 = \pi} = 3 K_2 K_4 - 4K_3^2 = B.$$
We already noted that $B<0$ (and hence $\nu_3 <0$) for coefficients defined in~\eqref{GCcoeff}. Thus, 
the NLS equation~\eqref{NLS} is defocusing. Steady states of~\eqref{NLS} have the form,
$$A(X,T) = \tilde{A}(X)e^{ i\kappa T},$$
where $\kappa\in\R$ and $\tilde{A}(X)$ satisfies the Duffing equation,
\begin{equation} \label{NLS_steady}
\pa_X^2 \tilde{A} = - \frac{1}{\nu_2}\left( |\kappa| \tilde{A} - |\nu_3| \tilde{A}^3    \right), 
\end{equation}
where $\kappa < 0$. 
Equation~\eqref{NLS_steady} possesses heteroclinic solutions 
which have the form,
$$\tilde{A}(X) = \sqrt{\kappa/\nu_3} \tanh\left( \sqrt{ \frac{-\kappa}{2 \nu_2 }}  (X- X_0 ) \right). $$
Reconstructing the original solution from our ansatz
of Eq.~\eqref{ansatz_strain}, we have the following approximation,
\begin{eqnarray*}\label{approx_james}
y_n(t) &\approx& \eps \tilde{A}e^{i\kappa T} e^{i(k_0n + \omega_0 t)} + \mathrm{c.c.} \\
  &=& 2 \eps (-1)^n   \sqrt{\frac{\kappa}{\nu_3}} \tanh \left( \sqrt{ \frac{-\kappa}{ 2 \nu_2 }}   \epsilon (n - x_0)   \right) \cos( \omega_b t )   \quad \quad \quad (\mathrm{for } \,\, k_0 = \pi)
\end{eqnarray*}
where $\omega_b = \omega_0 + \kappa \epsilon^2$ is the frequency of the breather, $\kappa <0$ is 
a fixed but arbitrary parameter and $X_0 = \epsilon x_0$ is an arbitrary spatial translation. When $\kappa < 0$, the frequency of the breather lies 
within the phonon band.
We note that the same approximation (up to first order) can be
obtained by performing a center manifold reduction (see e.g. \cite{James01,Rey04}
with $\kappa = -\nu_2$, $\eps = \sqrt{-\mu} $ and $A=M=1$ such that $\nu_2 = 1/4$).
It is relevant to mention that the defocusing NLS equation can also be derived in the displacement variable formulation \cite{Huang}.

If $\nu_3 > 0$ (which is not possible here), then one would need $\kappa>0$, which, in turn, would result in the existence of homoclinic solutions and thus 
of bright breathers with frequencies above the phonon band.

\section{The finite dimensional system and its conservation laws}
\label{numerics}

The boundary conditions (BCs) of the finite dimensional system used for the numerical simulations will play an
important role since the dark breathers have non-decaying tails.  In terms of the strain variable, a logical
choice is a periodic boundary condition $y_0 = y_N$ , $y_{N+1} = y_1$. 
To check our numerical results, we also carry out simulations
 in the displacement variables. 
The following system of $N+1$ equations is consistent with the system in the strain variable representation and 
periodic boundary conditions,
\begin{eqnarray*}
M \ddot{u}_1 &=& A  ( \delta_0 - ( u_{N+1} - u_{N} ))^{3/2}_{+}-A  ( \delta_0 - ( u_{2} - u_{1} ))^{3/2}_{+}, \\
M \ddot{u}_n &=& A  ( \delta_0 - ( u_{n} - u_{n-1} ))^{3/2}_{+}-A  ( \delta_0 - ( u_{n+1} - u_{n} ))^{3/2}_{+}, \quad n\in \{ 2,N\} \\
M \ddot{u}_{N+1} &=& A  ( \delta_0 - ( u_{N+1} - u_{N} ))^{3/2}_{+}-A  ( \delta_0 - ( u_{2} - u_{1} ))^{3/2}_{+},
\end{eqnarray*}
The redundancy of the last equation implies $\ddot{u}_{N+1} = \ddot{u}_{1}$, such that the above
may be written as a system of $N$ equations,
\begin{eqnarray*}
M \ddot{u}_1 &=& A  ( \delta_0 - ( u_1 + c_1 t + c_2 - u_{N} ))^{3/2}_{+}-A  ( \delta_0 - ( u_{2} - u_{1} ))^{3/2}_{+}, \\
M \ddot{u}_n &=& A  ( \delta_0 - ( u_{n} - u_{n-1} ))^{3/2}_{+}-A  ( \delta_0 - ( u_{n+1} - u_{n} ))^{3/2}_{+}, 
   \quad n\in \{ 2,N-1\}\\
M \ddot{u}_N &=& A  ( \delta_0 - ( u_{N} - u_{N-1} ))^{3/2}_{+}-A  ( \delta_0 - ( 
u_1 + c_1 t + c_2  - u_{N} ))^{3/2}_{+},
\end{eqnarray*}
where  $c_1 = \dot{u}_{N+1} (0)- \dot{u}_{1}(0)$ and  $c_2 = {u}_{N+1} (0)- {u}_{1}(0)$.  This system 
maintains the conserved quantities of the infinite system. For example, if $c_1=0$, then
the Hamiltonian and mechanical momentum,
$$ H = \sum_{n=1}^{N-1} \left[ \frac{1}{2} M v_n^2 + V_{\mathrm{GC}}(u_{n+1}-u_{n}) \right] + V_{\mathrm{GC}}(u_1 + c_2 -u_{N}),  \quad \quad P = \sum_{n=1}^{N} M v_n,$$
are conserved respectively.
The conservation of the Hamiltonian is connected to the temporal translation invariance $t \rightarrow t + \delta t$ and the conservation of momentum is a result of the vertical translation invariance $u = u + \delta u$.
An interesting feature 
is the invariance,
\begin{equation} \label{shiftvar_dis}
 u_n \rightarrow \alpha u_n + c n, \quad \quad t \rightarrow \alpha^{-1/4} t , \quad \quad \alpha = (\delta_0 - c)/\delta_0,
 \end{equation}
which holds in the infinite and finite lattice using the above mentioned boundary conditions. 
This last invariance is connected with the existence of a marginal mode and it has important consequences
 in the linear stability analysis described in Sec.~\ref{LSA}.

\section{Bifurcations}
\label{bifurcations}

\begin{figure} 
 \centerline{ 
\epsfig{file=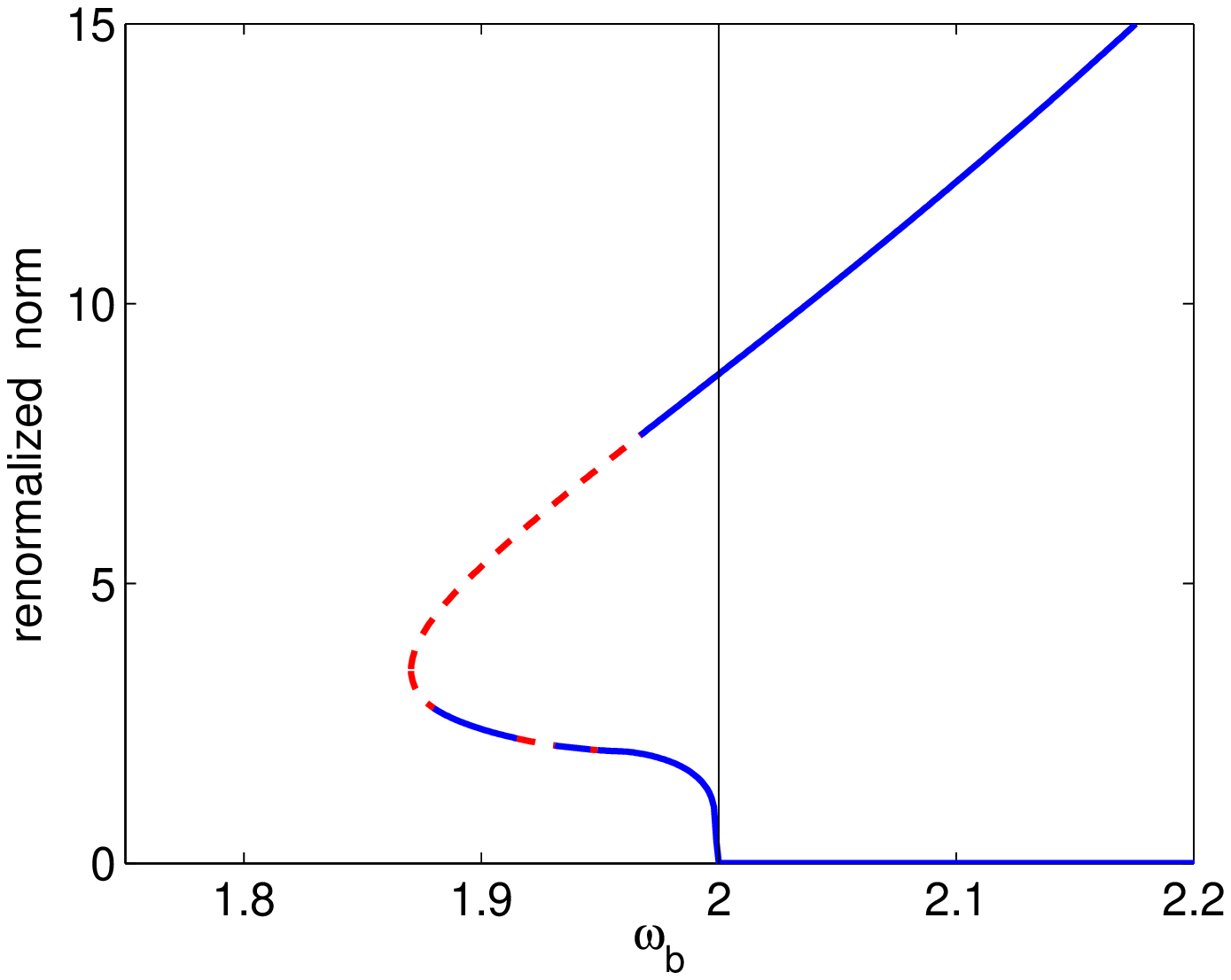,width=.45 \textwidth}
\epsfig{file=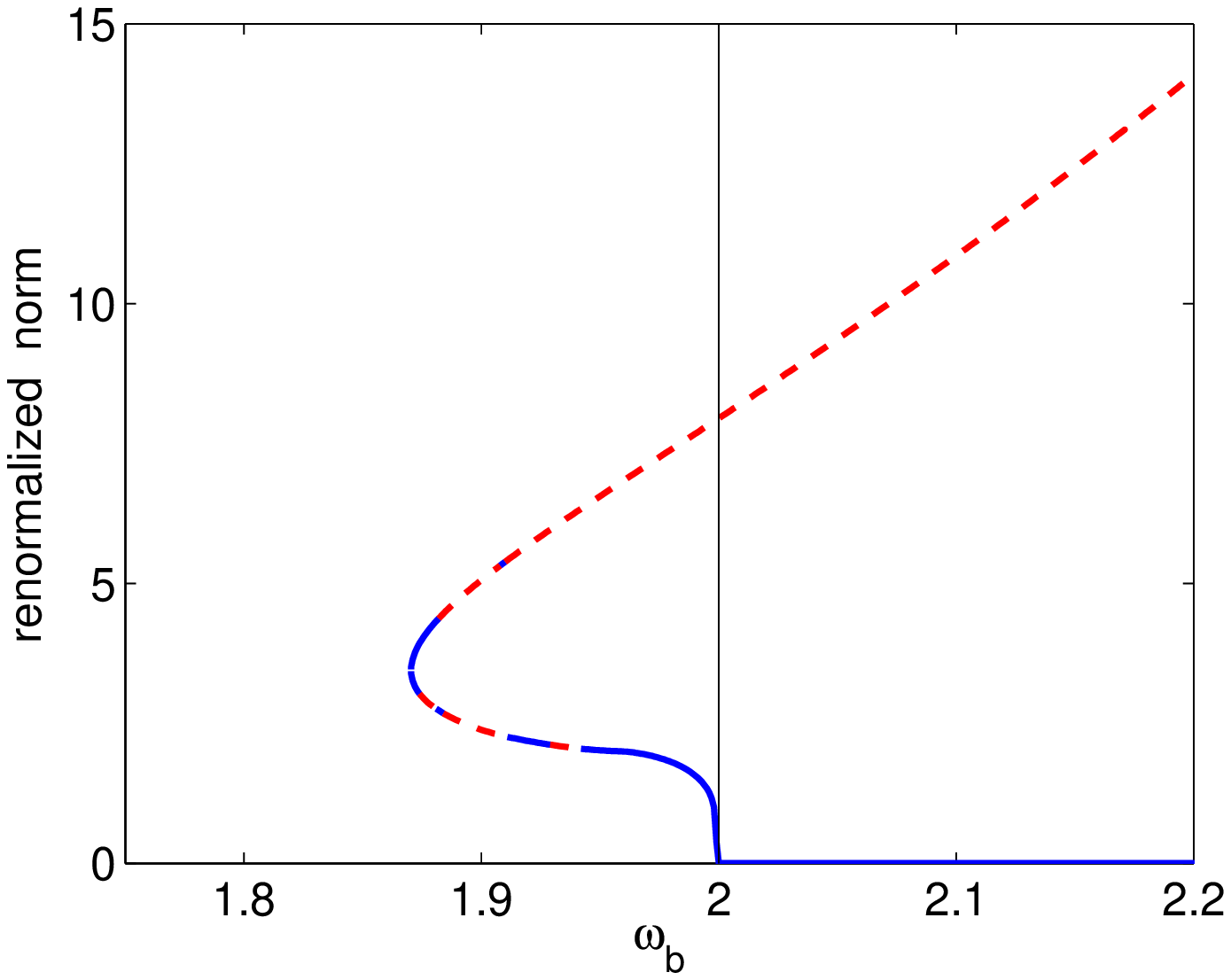,width=.45 \textwidth}
 }
\caption{ Left:  Renormalized $l^2$ norm of the site-centered solution versus $\omega_b$ with a vertical center of  $c=0$. 
The vertical line shows the edge of the phonon band $\omega=2$.  The upper branch of the curve reaches past the edge of the
phonon band and increases indefinitely. Right: Same as left, but for the bond-centered solution. Regions where a real instability is present are indicated by a dashed red line (see Sec.~\ref{LSA} for details).}
\label{bifurcation}
\end{figure}

We compute numerically exact dark breather solutions using a Newton-type solver for time-periodic solutions (see e.g. \cite{Flach2007} for details).  We carry out the computations in the strain variables with periodic boundary conditions with a lattice size of $N=51$. The associated Jacobian matrix
is rank deficient due to the invariances of the system, and therefore
an additional constraint is needed. It is sufficient to fix the vertical center of the solution,
 $$c := \frac{\sup_{t \in [0,T_b]} y_{1}(t) + \inf_{t \in [0,T_b]} y_{1}(t)}{2}, $$
%
where $T_b = 2\pi/\omega_b$ is the breather period. For all reported results
we used the parameter values $M=\delta_0=1$ and $A=2/3$ such that $\omega_0=2$. After a rescaling of time and amplitude, solutions for arbitrary parameter values can be recovered.
The $l^2$ norm and energy of the dark breathers will depend on the nonzero background of the solutions. 
Thus, to measure the solutions we consider the renormalization \cite{Kivshar94},
$$ \tilde{y}_n = \mu - | y_n - c|  \quad \quad \quad   \| y_n \| _{\tilde{l}^2} = \sum_{n \in I} \mu^2 - |y_n - c |^2, $$
where $\mu$ is defined as the amplitude of the breather,
$$ \mu := \frac{\sup_{t \in [0,T_b]} y_1(t) - \inf_{t \in [0,T_b]} y_1(t)}{2}. $$
The renormalized profiles can be seen in the middle panels of Fig.~\ref{profiles}. We
use the analytical approximation~\eqref{ansatz_strain} to seed the numerical solver.
Ansatz~\eqref{ansatz_strain} with $x_0=0$ represents a so-called site-centered solution (see Fig.~\ref{profiles} top) whereas the bond-centered
solution corresponds to $x_0 = 0.5$ (see Fig.~\ref{profiles} bottom). 

In the finite lattice, eigenvectors of the equations of motion linearized about the trivial solution also provide an option for an initial seed.
The largest computed eigenvalue $\omega$ of the linear spectrum
serves as the numerical cutoff value of the frequency (which is slightly 
smaller than the exact value of
$\omega_0$ but approaches it as $N\rightarrow \infty$). The eigenvector of the $N-1$ eigenvalue has the profile of a dark breather.
The drawback of using the linearized solution is that there is no information 
regarding the correct amplitude. For this reason,
we use~\eqref{ansatz_strain} for the initial seed, which approximates both the shape and amplitude 
of the solution. 

In the left panel of Fig.~\ref{bifurcation}, we show
 the renormalized $l^2$ norm of the numerically exact site-centered dark
breathers. The solutions do not exist for arbitrarily small $\omega_b$
but rather, there is a fold at $\omega_b \approx 1.87$. 
However, we note here that this is a fold in the particular parameters
which does not constitute a point of change in the spectral stability
of the observed states.
The top branch  
persists past the edge of the phonon band $\omega_0 =2$,
with indefinitely increasing amplitude. For this reason, the solutions making up the top branch
in the figure can be thought of as strongly nonlinear solutions, and those making up the
bottom branch as weakly nonlinear solutions. A red dashed line indicates that
the solution possesses a real instability, a blue solid line indicates the solution does not (see Sec.~\ref{LSA} for more details).

\section{Dark breathers with a nonzero center}

\begin{figure} 
 \centerline{ 
\epsfig{file=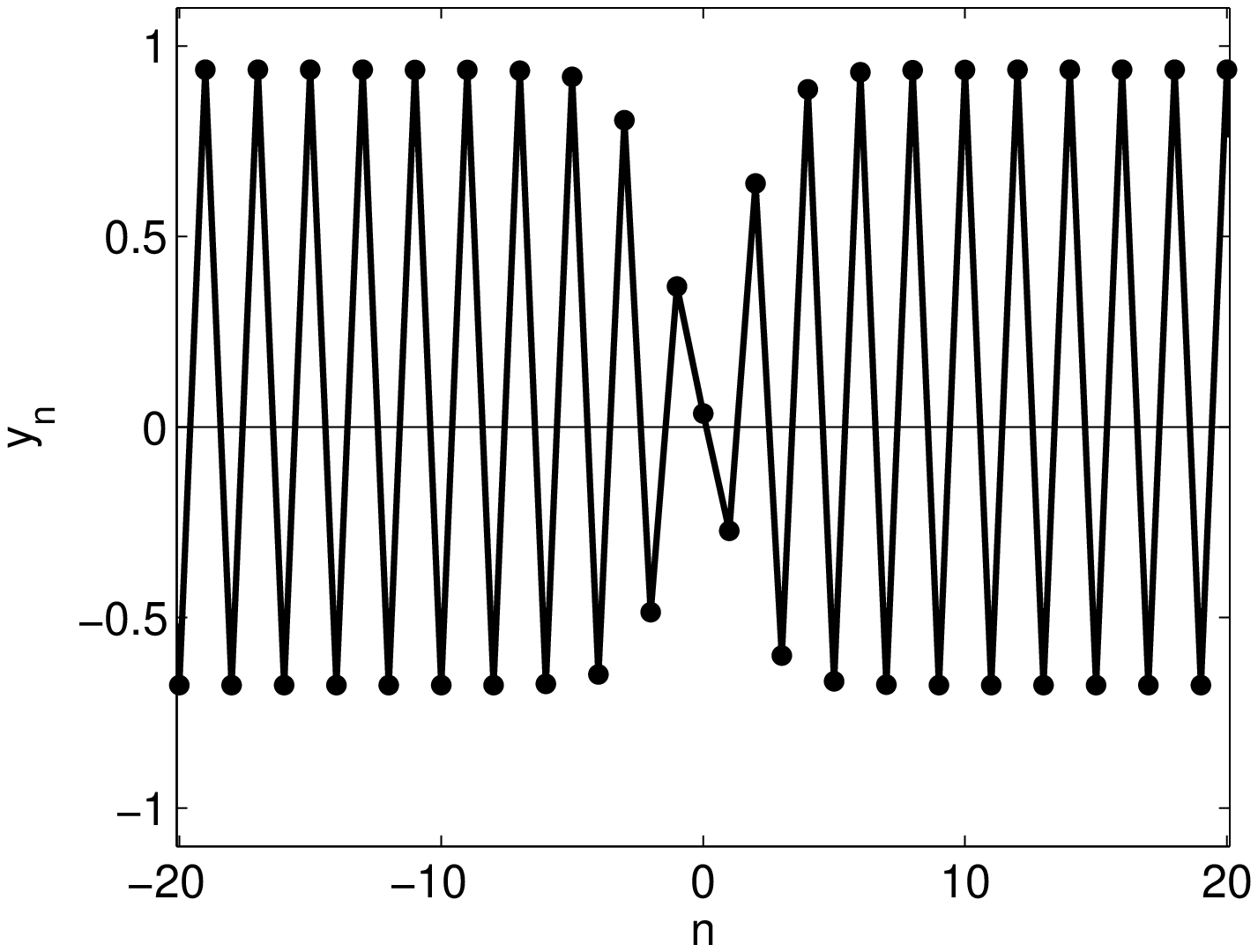,width= .3 \textwidth}
\epsfig{file=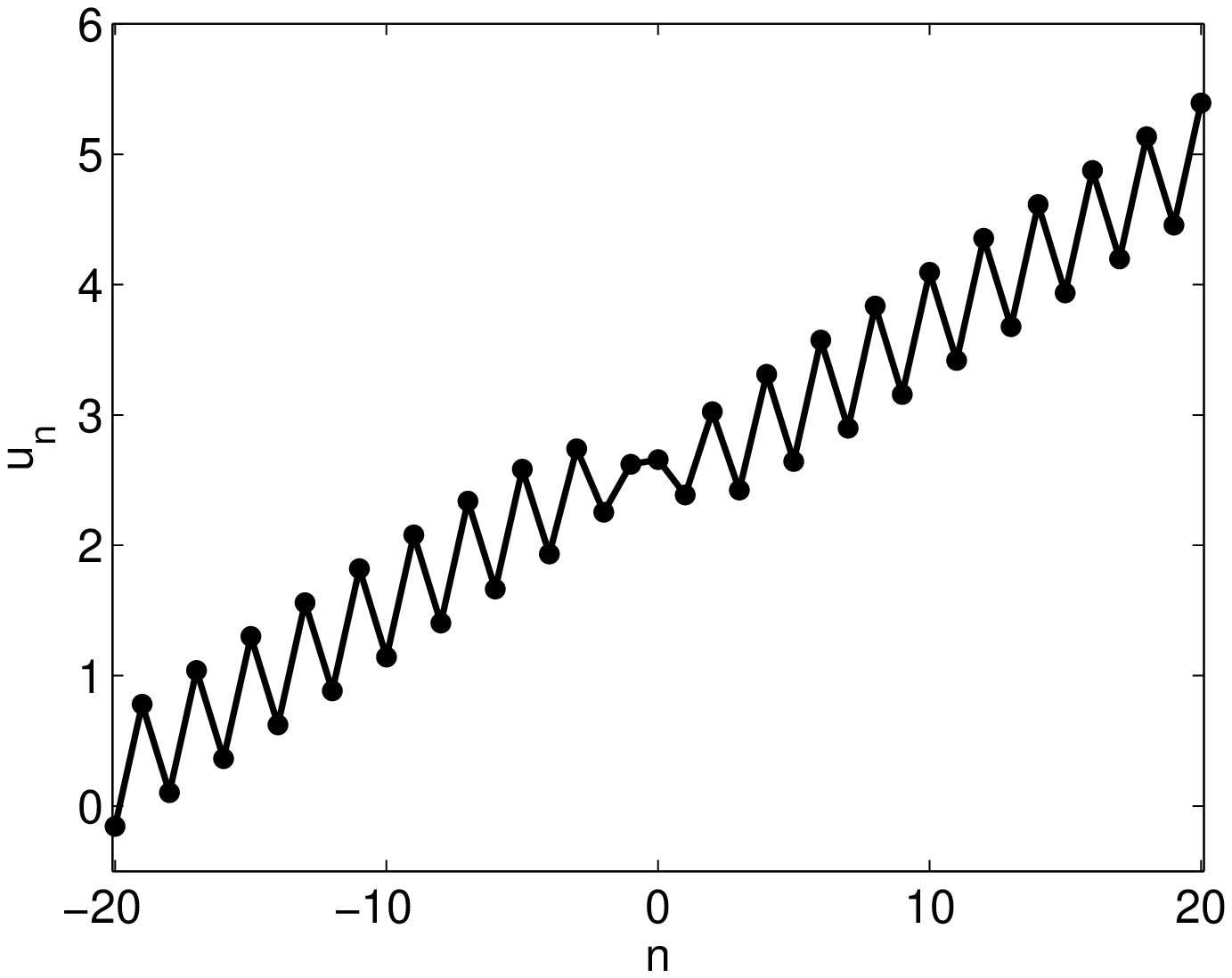,width=.3 \textwidth}
\epsfig{file=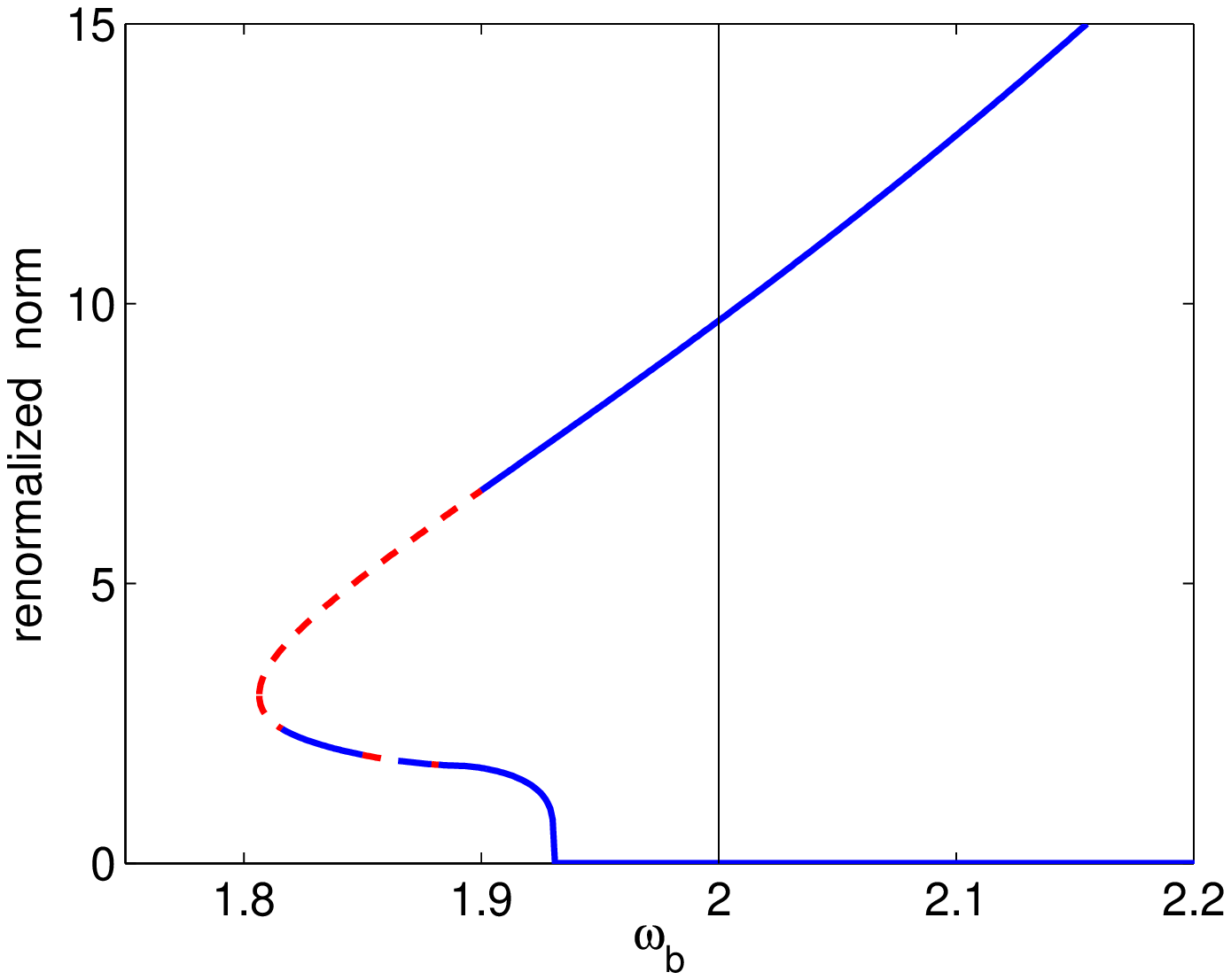,width=.3 \textwidth}
}
\caption{Left: A site-centered dark breather solution with $\omega_b = 1.9$ and a nonzero center $c=0.13$ in the strain variable $y_n$.
Middle: The same solution but in the displacement variables $u_n$.
Right:  Renormalized $l^2$ norm of the 
site-centered solution versus $\omega_b$ with a vertical center of $c=0.13$. 
The vertical line shows the edge of the phonon band $\omega=2$. In contrast to the $c=0$ case shown in Fig.~\ref{bifurcation}, the bottom
branch terminates at a value below the edge of the phonon band. }
\label{bifurcation_c13}
\end{figure}

The equations of motion for both the infinite system and the finite system with periodic BCs are invariant under
the transformation
\begin{equation} \label{shiftvar_strain}
y_n \rightarrow y_n \alpha + c, \quad \quad  t \rightarrow \alpha^{-1/4}t, \quad \quad \alpha = (\delta_0-c)/\delta_0,
\end{equation}
where~\eqref{shiftvar_strain} is merely the strain variable equivalent of \eqref{shiftvar_dis}. 
Thus, there is an entire family of solutions associated with the vertical shift $c$. 
We note that a similar transformation has been introduced in \cite{Rey04} for the case of FPU lattices.
Figure~\ref{bifurcation_c13} shows an example of a dark breather with
$c=0.13$ in the strain (left) and displacement variables (middle).
The bifurcation diagram (right) 
is similar to its $c=0$ counterpart, but is shifted in the $\omega$ axis  where the cutoff point is $\omega_{0} \alpha^{1/4}=2 \alpha^{1/4} $. The invariance~\eqref{shiftvar_strain} can also be used as
another accuracy check for the numerical simulations. We can choose the value of the constraint $c$ in the Newton method and
test if it converges to the appropriate solution as predicted by~\eqref{shiftvar_strain}, assuming we have computed
the $c=0$ solution. Naturally, the stability of the two solutions
also has the same structural characteristics, as regards intervals
with real Floquet multipliers.

Note that a constant vertical shift $y \rightarrow y +c$ is the same as leaving $y$ unchanged and changing the precompression
$\delta_0 \rightarrow \delta_0 - c $. Thus, all forthcoming discussions of changes in vertical center can be thought of as changes in
precompression. For this reason it is clear that both the strongly and weakly nonlinear branch go toward the zero solution as the precompression
goes to zero, since $\alpha \rightarrow 0$ as $c \rightarrow \delta_0$.

\section{Linear stability analysis}
\label{LSA}

\begin{figure} 
 \centerline{
\epsfig{file=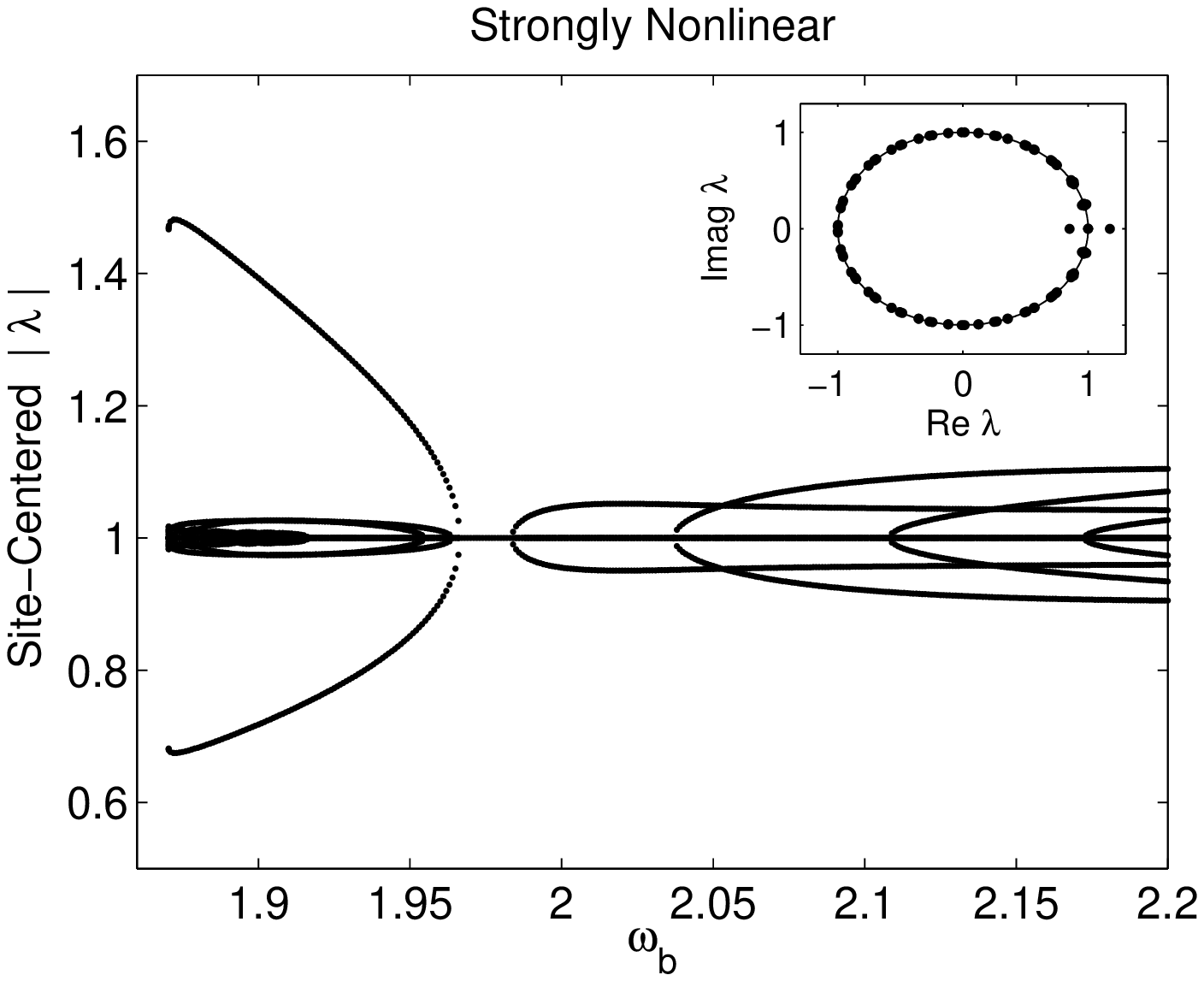,width= .45 \textwidth}
\epsfig{file=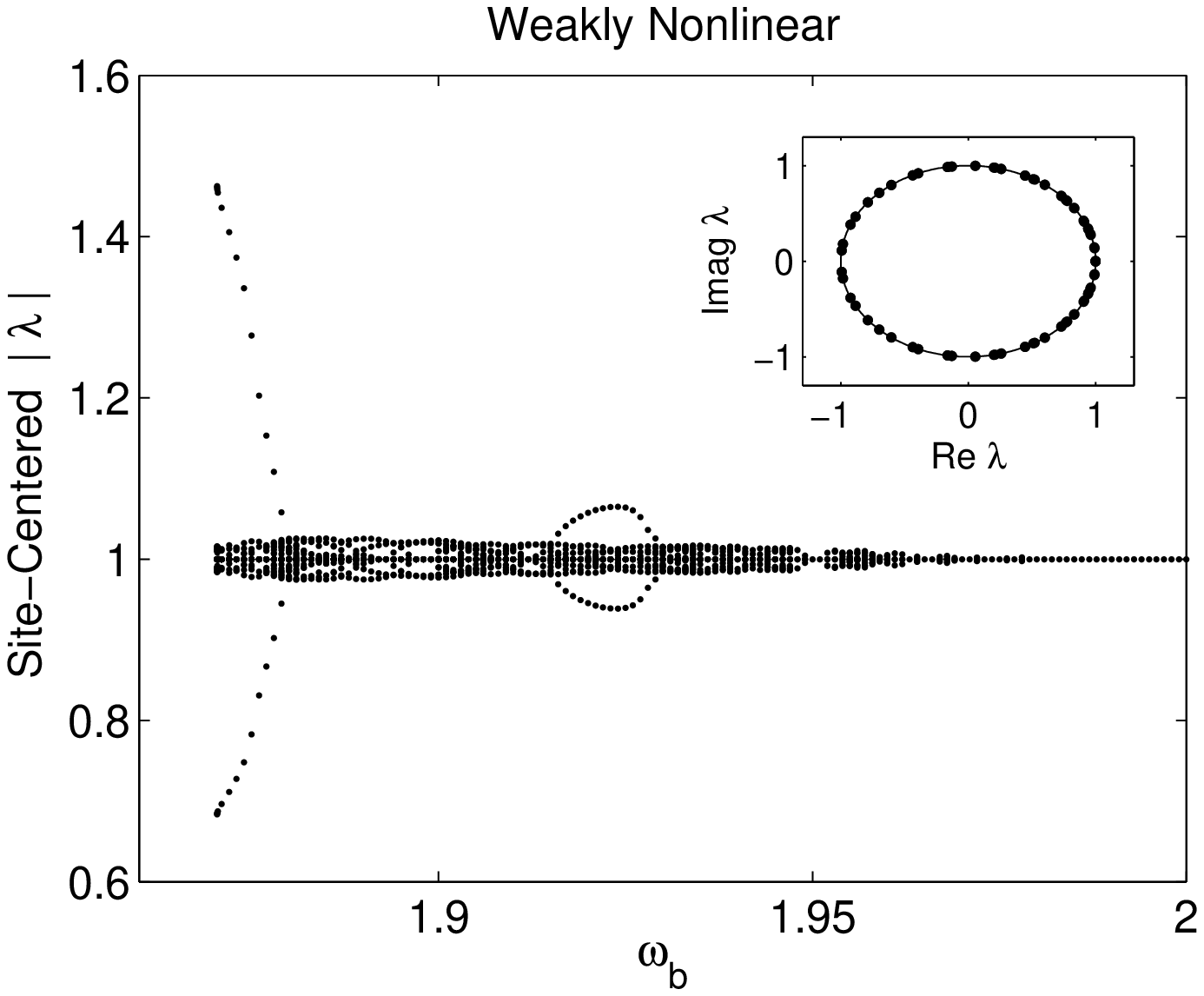,width= .45 \textwidth}}
\centerline{
\epsfig{file=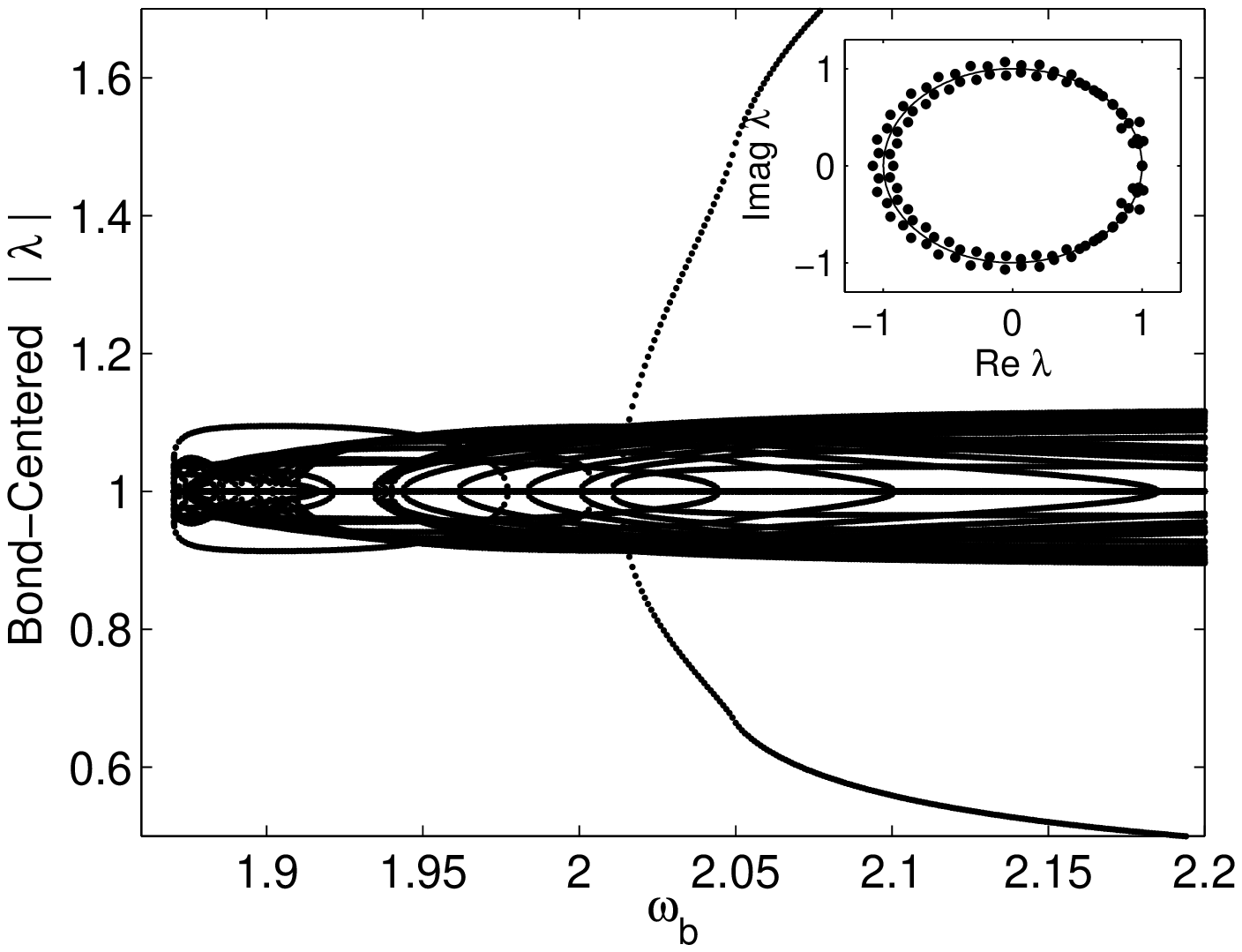,width=.45 \textwidth}
\epsfig{file=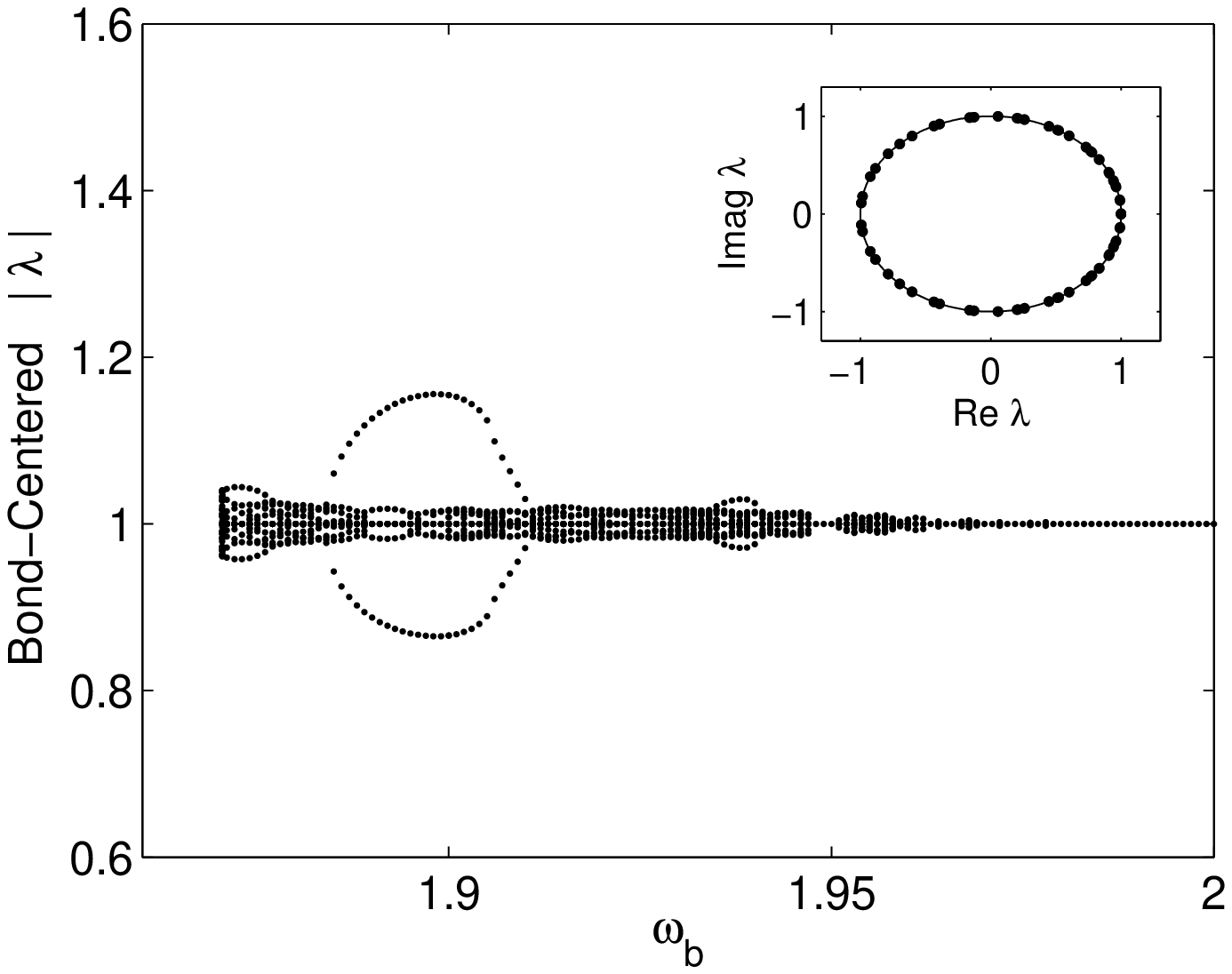,width=.45 \textwidth}
 }
\caption{Floquet multipliers of the strongly (left) and weakly (right) nonlinear 
solutions with $c=0$  versus frequency $\omega$ for the site-centered (top) and
bond-centered (bottom) types. The Floquet multipliers for $\omega_b=1.95$ in
the complex plane are shown in the respective insets.}
\label{stab}
\end{figure}

There is a rich theory for the stability properties of breathers 
in Hamiltonian systems \cite{Flach2007,Aubry06,Flach05}. Here, linear stability is determined in the standard way:
A perturbation $V_n(t)$ is added to a breather solution and  \eqref{GC_strain}
is used to derive an equation describing the evolution of $V$. Keeping 
only linear terms
in $V$ will result in a Hills' equation with a time periodic coefficient of period $T_b$. Thus, 
the eigenvalues (or Floquet multipliers) of the associated 
variational matrix at $t=T_b$ determine the linear stability of the breather 
solution. Due to the
Hamiltonian structure of the system, all Floquet multipliers must lie on the unit circle for the solution to be (marginally)
stable, otherwise, the solution is unstable.

There are continuous arcs of spectrum on the unit circle (in the infinite
lattice limit), which can be computed from
the phonon band of~\eqref{GC_strain}. Although, in contrast to the bright 
breather case,
these arcs are not simply $ \exp( i \omega \, T_b ) $, where $\omega \in [-\omega_0 , \omega_0 ] $.
This is due to the nonzero background of the solutions we are linearizing about.
The isolated (point spectrum) multipliers must be computed numerically. If an 
isolated one is real and lies off the unit circle, then the solution possess
a so-called real-instability  (see Fig.~\ref{stab}, top left for an 
example).  Oscillatory instabilities are also possible, 
where the multiplier  has both real and imaginary parts (and the corresponding
multipliers come in quartets). In FPU lattices, it is known that numerically
computed eigenvalues may lie off the unit circle due to the truncation of
the infinite lattice to a finite lattice \cite{Flach2007,Aubry06}. To determine if the instability
is a finite size effect, a band analysis can be carried out \cite{Flach2007}. 
Such a computation
is outside the scope of this work, but we did carry out numerical simulations 
for larger values of 
the lattice size $N$. We found that the oscillatory multipliers do approach 
the unit circle, but 
level off before reaching it. In other words, the oscillatory instabilities 
seem to be genuine, albeit weak.

\begin{figure} 
 \centerline{
\epsfig{file=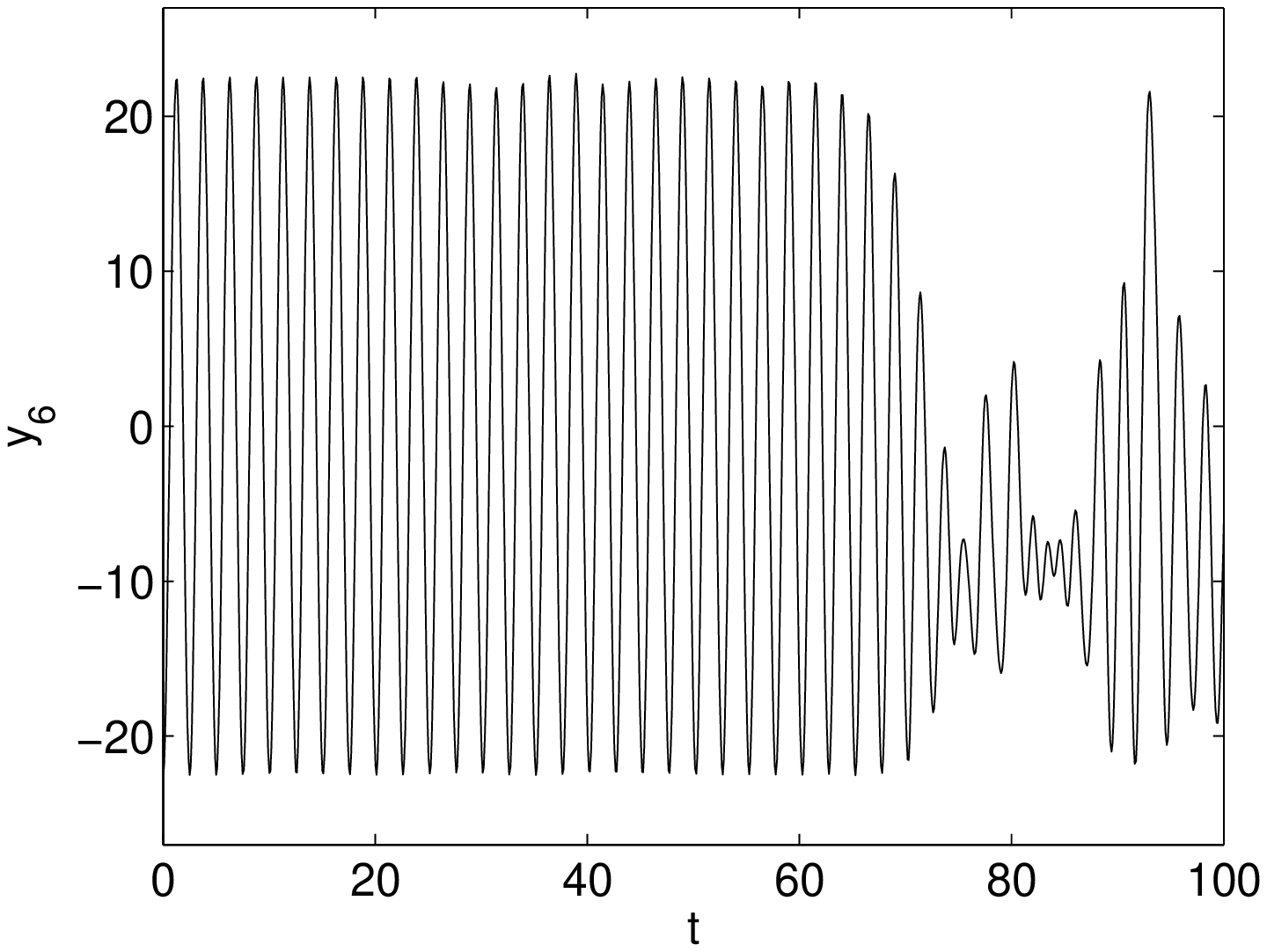,width=.45 \textwidth}
\epsfig{file=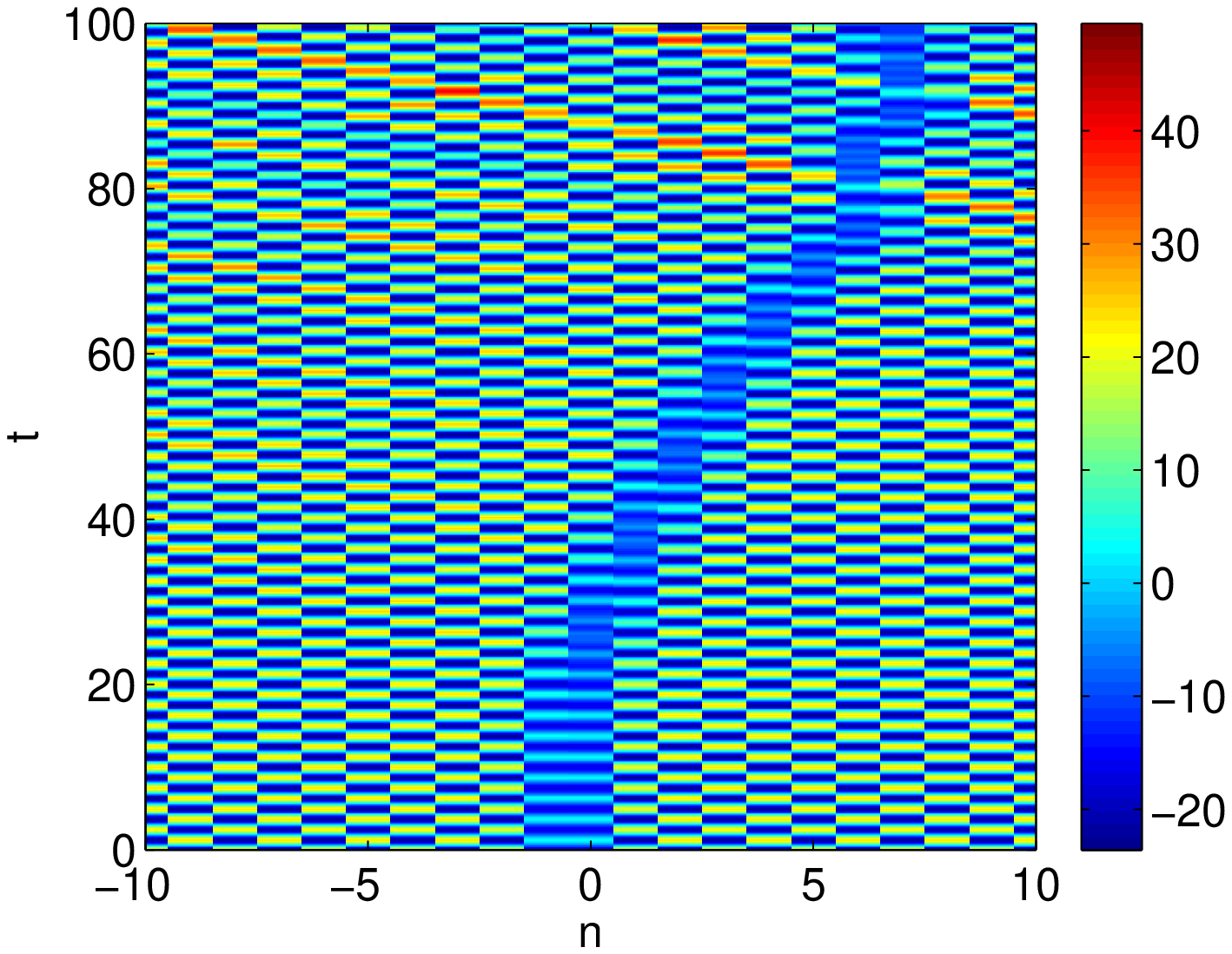,width=.45 \textwidth}}
\caption{Manifestation of the real-instability of the strongly nonlinear bond-centered solution for $\omega_b = 2.5$.
The left panel shows the evolution of the $n=6$ node. The right panel is a contour plot of the time evolution of
the entire solution. The color intensity corresponds to the value of the strain $y_n$. The checkered pattern is a
consequence of the time periodicity of the solution. Notice the density dip of the solution near $n=0$, and how this structure re-arranges itself
in a way reminiscent of a moving (dark breather) pattern after $t \approx 25$. }
\label{instability}
\end{figure}

Examples of the numerically computed Floquet spectrum of the site-centered and bond-centered
solutions, for both strongly and weakly nonlinear types, are shown in the insets of Fig.~\ref{stab}
for $\omega_b = 1.95$. The moduli of the Floquet multipliers are also shown for various frequencies where real and oscillatory instabilities are present.  The real instabilities are the larger arcs in the figure. For each solution type, 
there
is an intricate cascade of oscillatory instabilities, which can be best seen
for the strongly nonlinear site-centered solution. 
The magnitude of the oscillatory instabilities is small relative to that of the
real instabilities.  Therefore, in the bifurcation diagram of Fig.~\ref{bifurcation}
only real instabilities are indicated, although  oscillatory
instabilities may be present. In Fig.~\ref{instability} the manifestation of an instability of the strongly 
nonlinear bond-centered solution for $\omega_b=2.5$ is shown, which is characteristic 
of the simulations carried out for the perturbed unstable solutions. 
It can be clearly seen that the instability essentially sets the
dark breather state in motion, although it 
does not appear to correspond to a genuinely traveling breather.

The linear stability analysis (LSA) described above can fail to describe instabilities that grow slower than exponentially \cite{Aubry06}. 
Such "nonlinear" instabilities can emerge even if the linear stability analysis predicts the solution to be marginally stable. They are associated with degenerate 
eigenvalues (i.e. unit Floquet multipliers),
although the presence of degenerate eigenvalues does not, in turn, 
guarantee the 
existence of a nonlinear instability.
For solutions with {\it $|y_n| < \delta_0$} there are, up to machine precision, six 
degenerate eigenvalues. One pair, corresponding to the phase mode $V_p =\{ \dot{y}(0), \ddot{y}(0) \}$, 
is associated with the energy conservation/time reversal invariance of the system. The second pair associated to a pair of 
degenerate eigenvalues seems to be of the form $V_t = \{  \pa y/ \pa {x_0} , \pa \dot{y}/ \pa {x_0} \} $ which corresponds 
to continuous spatial translation invariance. This mode, called also "translational" or "pinning" mode, is of particular interest 
since it suggests that (exact) traveling dark breathers may
exist in this model. Indeed for frequencies very close to the band edge, we found several intermediate solutions, which
seem to be spatial translations of the same solution (i.e. solutions with various values of $x_0$ in Eq.~\eqref{ansatz_strain}).
For solutions with $|y_n| \geq \delta_0$ this invariance is lost, and there are only four degenerate eigenvalues
in the linear stability problem.

The last pair of modes has the form
$V_c = \{  \pa y/ \pa {c} , \pa \dot{y}/ \pa {c} \}$
%
where $c$ is the vertical center. To the best of the authors knowledge, these modes have never been 
found/examined in FPU lattices. Perturbing along the direction of these modes results in algebraic growth in the
perturbation. Thus, we conjecture that there is a nonlinear instability associated 
with this marginal mode, (which in turn is connected to the shift 
invariance~\eqref{shiftvar_strain}). The heuristic argument for the 
presence of such an instability
is as follows: if one chooses a small perturbation of the form $\epsilon_n + c$, the dynamics of the solution with a vertical center of $c$
may emerge, which will have a lower frequency, due to~\eqref{shiftvar_strain}. 
This is precisely
what we observe in the numerical simulations with such perturbations. For example, the weakly nonlinear
site-centered solution at $\omega_b=1.97$ is linearly stable. However 
a perturbation with initial value $V_n(0) = .01 \max( y(0) ) ( r_n +  c) $,
where $r_n$ is a random perturbation centered at 0 with $|r_n| < 1$, will grow even though the LSA predicts that it should not.  
Moreover, the projection 
$p(t) = <y_{\mathrm{exact}}(\cdot , t) - y_{\mathrm{perturbed}}(\cdot,t), V_c> $  grows algebraically
and is several orders of magnitude greater than the 
projections to the other linearization eigenvectors. This
phenomenon is not observed when the above perturbation has $c=0$
for the identical timescale.
Lastly, 
if the invariance~\eqref{shiftvar_strain} is absent, the nonlinear 
instability vanishes (see Sec.~\ref{fixedbc} for a relevant example).


We note in passing, that working on the equivalent displacement representation, 
see Sec.4, a fourth pair of Floquet multipliers is located at $(1,0)$ in the complex plane. 
This pair arises from the conservation
of the total mechanical momentum.


\section{Other boundary conditions}
\label{fixedbc}

\begin{figure} 
 \centerline{ 
\epsfig{file=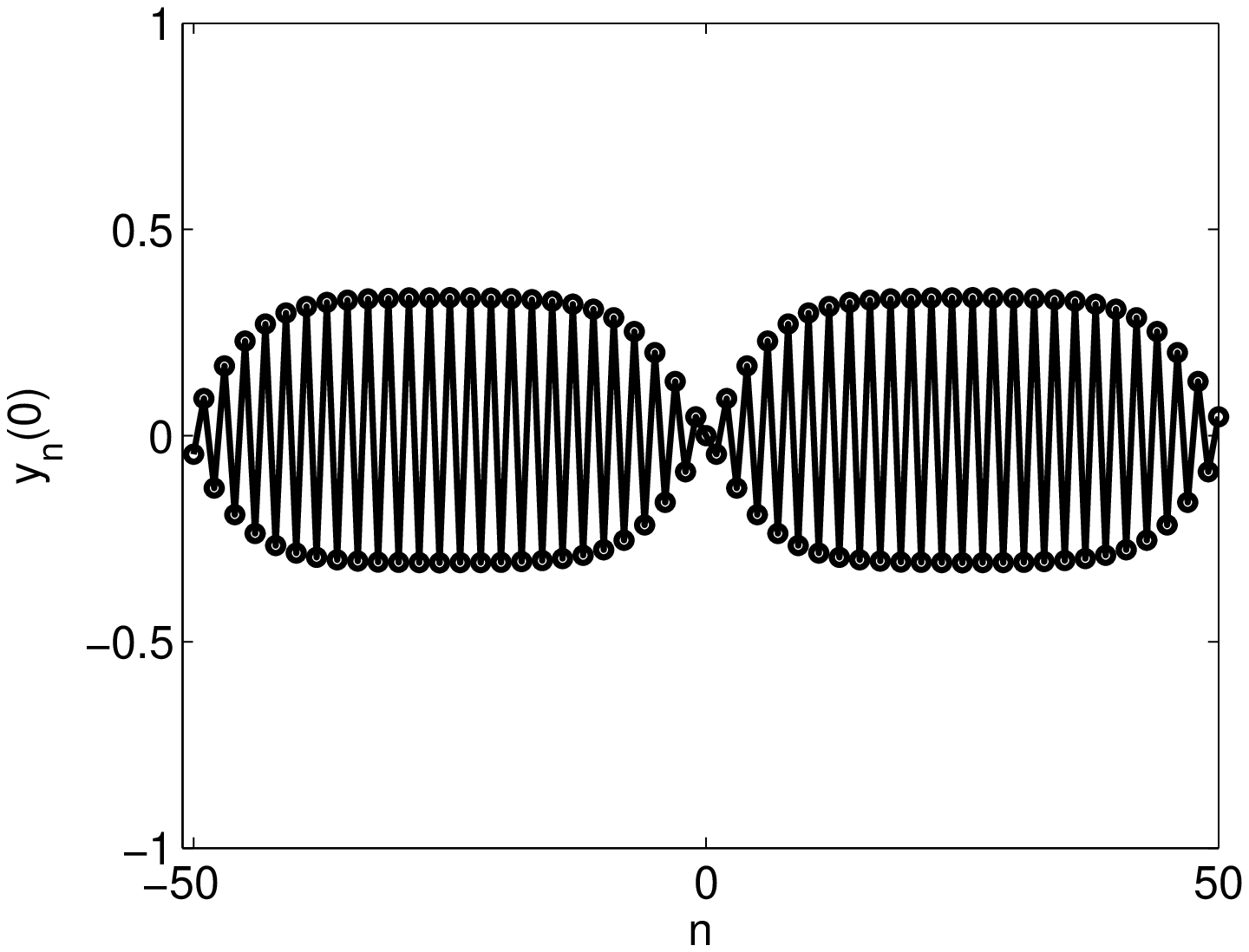,width= .45 \textwidth}
\epsfig{file=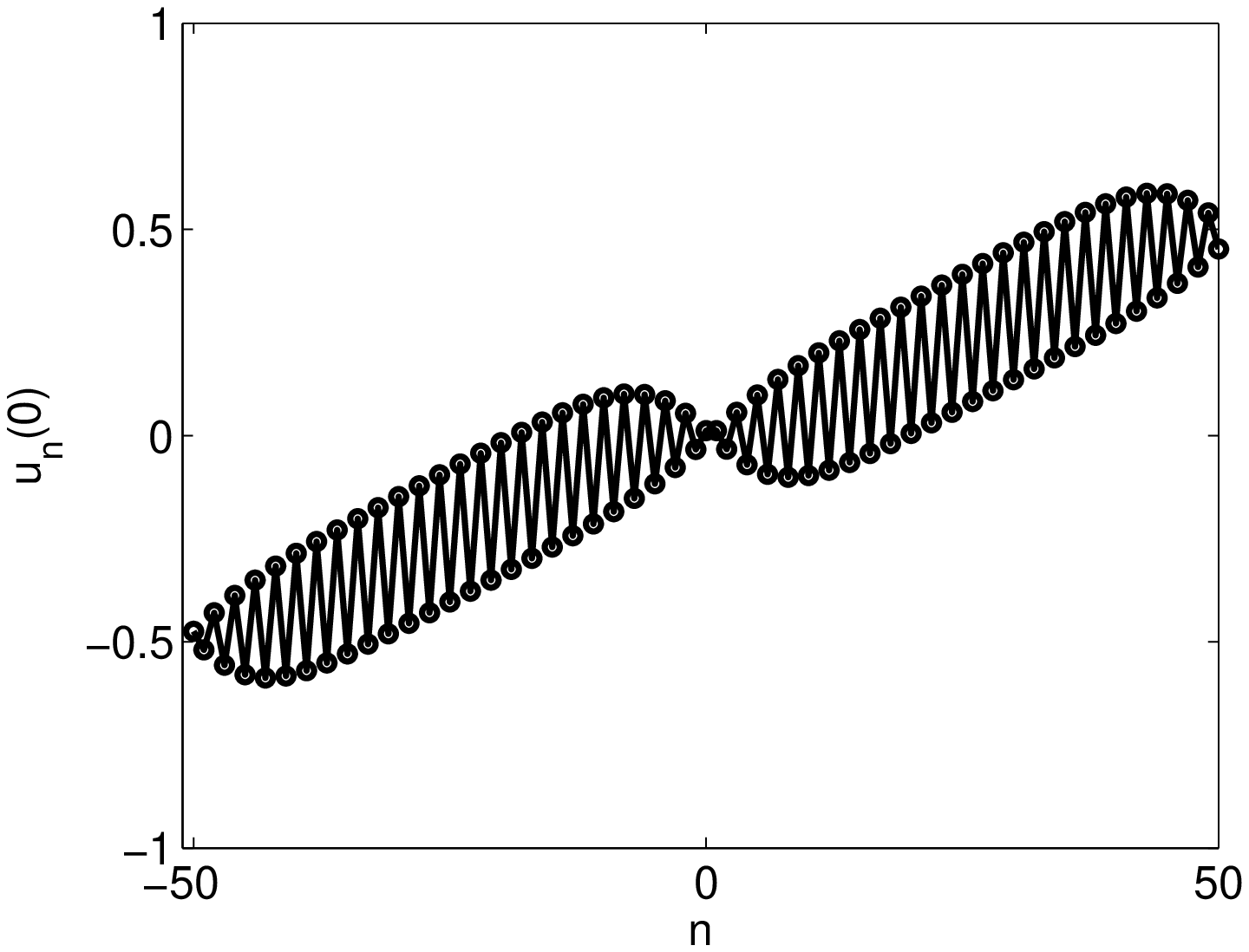,width=.45 \textwidth}
}
\caption{Left: A site-centered dark breather solution with $\omega_b = 1.99$ in the strain variable $y_n$ with fixed boundary conditions $y_0=y_{N+1}=0$ with $N=101$.
Right: The same solution but in the displacement variables $u_n$.
 }
\label{fixed_profiles}
\end{figure}

Periodic boundary conditions are advantageous since the solution profiles 
closely resemble those of the infinite system, and invariances of the 
infinite system carry over to the finite case. 
However, other boundary conditions are also relevant and interesting to
examine
from an experimental point of view.  The free boundary
conditions $u_0 := u_1$ and $u_{N+1} := u_N$ constitute such an
example. In terms of strain variables, these become 
fixed boundary conditions $y_0 = y_{N+1} = 0$.
The resulting profile resembles a multi-breather, see Fig.~\ref{fixed_profiles}. This solution is no longer centered at zero.
Unlike the periodic BC situation, one
cannot normalize the solution to have a zero center, since the vertical
shift invariance~\eqref{shiftvar_strain} is broken in the fixed BC system. Consequently, the corresponding marginal mode discussed in~Sec.\ref{LSA}
vanishes, along with the eigenvector $V_c$ and hence the nonlinear 
instability is absent in this case. Thus, choosing a strictly positive perturbation will no longer cause the integrator to shift towards a solution
of a different frequency, since the corresponding family of vertically shifted solutions no longer exists. Indeed, in the simulations a perturbation of the form
$\epsilon_n +c$ induces the nonlinear instability in the periodic BC system, 
but the \emph{exact same} perturbation in the fixed BC system
does not lead to any instability and the corresponding dark breather is
found to be structurally robust (where appropriate), 
in agreement with the perturbations
of the linear stability analysis presented above.

\section{Accuracy of the NLS approximation}

\begin{figure} 
 \centerline{ 
\epsfig{file=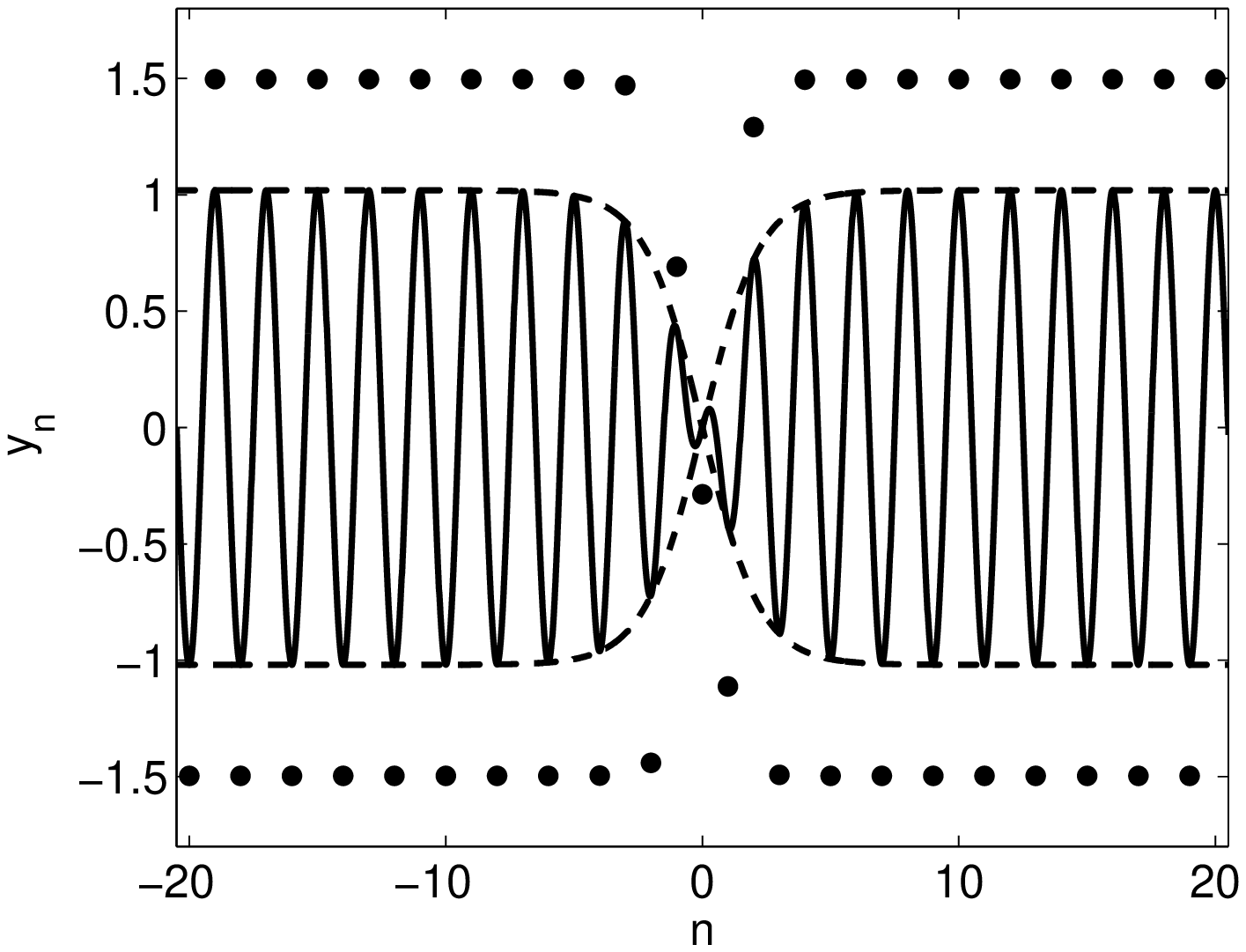,width=.45 \textwidth} 
\epsfig{file=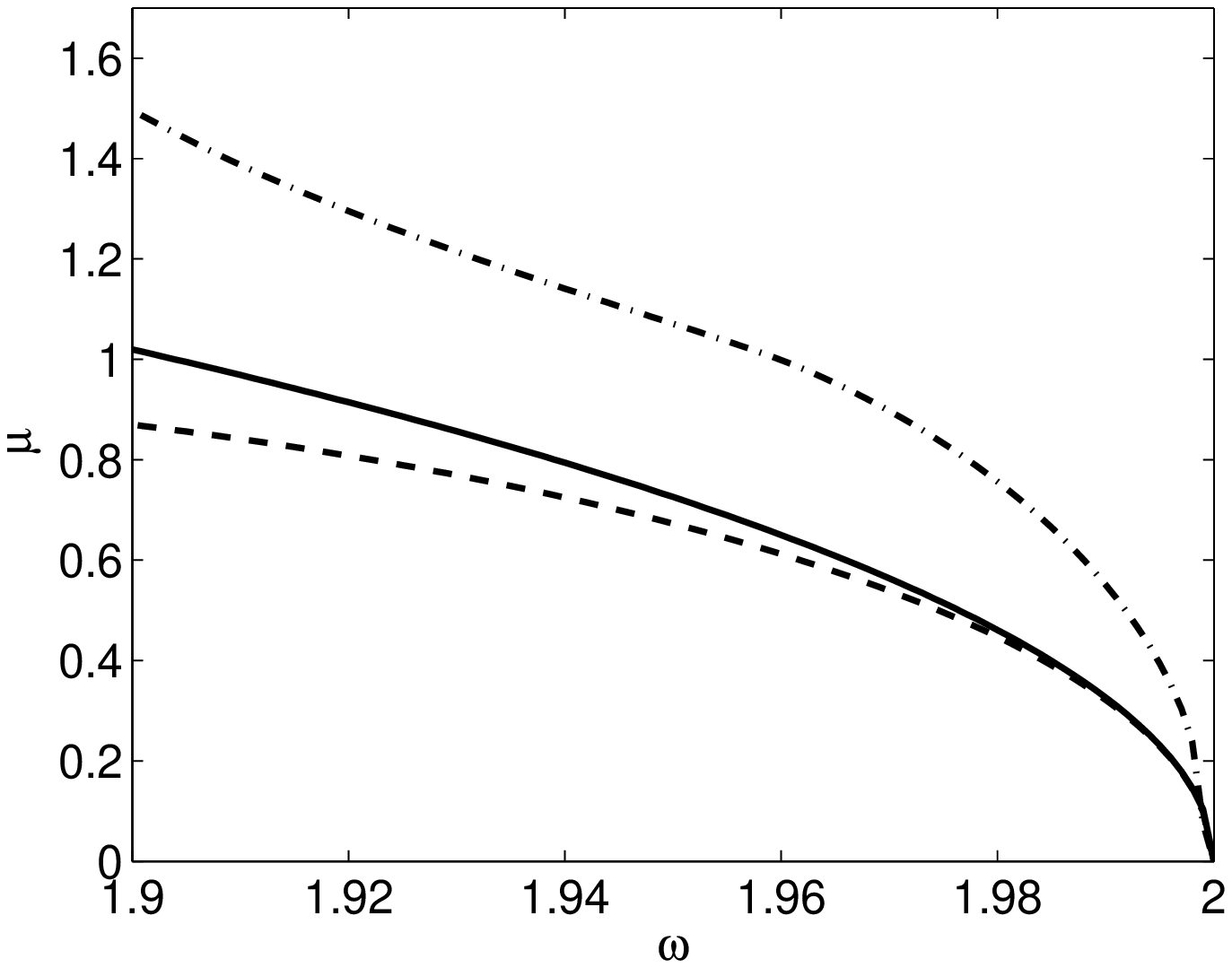,width=.45 \textwidth}  
 } 
\caption{Left: Analytical approximation (solid line) with an envelope function that is the NLS dark soliton (dashed line),
and a numerically exact dark breather (markers) for $\omega_b=1.9$ and $c=0$.  Right: The amplitude of the NLS approximation (solid line), numerical breather
with fixed $c=0$ (dashed-dot line) and numerical breather with varying $c$ (dashed line).    }
\label{validity}
\end{figure}

Although the NLS approximation~\eqref{approx_james} was sufficiently accurate to provide an initial seed for our numerical solver,
it is relevant to investigate the extent to which the NLS approximation 
represents actual solutions to~\eqref{GC_strain}.
Figure~\ref{validity} shows a numerically exact solution in the strain
variable (markers), the corresponding approximation (solid line) and the envelope (dashed line), which is defined by the NLS equation. 
From a visual inspection, the structure is quite proximal to the
exact one, although the amplitude
is underestimated. We also directly
compared the amplitudes of the analytical prediction with the numerical solutions (right panel) for various $\omega_b$, 
 where it can be seen that NLS approximation becomes irrelevant as $\omega_b$ moves away from the cutoff point $\omega=2$. 
It can perhaps be understood at an intuitive level, by inspecting 
Figure~\ref{validity}, that the NLS approximation is likely to be less
successful for the dark breather states considered herein than the more
standard case of the bright breathers~\cite{Flach2007}. In the latter
case, the solutions bifurcate from vanishing amplitude, on top of a 
vanishing background, while for the solutions considered herein
 the amplitude of the background is
finite even close to the to the cutoff limit. Hence, the cubic 
nonlinearity approximation is expected to be less adequate even in the
vicinity of that limit. 
Moreover, the fold bifurcation is not predicted by the NLS approximation 
and there is no approximation for the strongly nonlinear modes. 
Thus, the NLS approximation is only relevant for the weakly
nonlinear solutions near the edge of the phonon band, as expected. 
It is interesting to point out that the 
amplitude of the dark breather with fixed
$c=0$ is not the closest one to that of the NLS approximation.

The NLS approximation has been rigorously justified in the
context of the FPU lattice in~\cite{Schn10}, and more recently for an FPU 
equation with an arbitrary periodic configuration
in \cite{CCPS12}. There, it can be shown that solutions with initial data defined by the NLS equation will remain close
to an actual solution on long, but finite time scales, where the wavenumber $k$ can be arbitrarily chosen.
However, those results do not apply to dark breathers, the issue being,
as highlighted also above, the nonzero background.
The results presented in this work suggest that the justification 
considered in these works may be an interesting topic to investigate
in the context of granular crystals, but also perhaps more generally
(even for smoother potentials) for dark breather states existing on
a finite amplitude background.

\section{Exciting the dark breathers and dissipative effects}
\label{diss}

\begin{figure} 
 \centerline{\epsfig{file=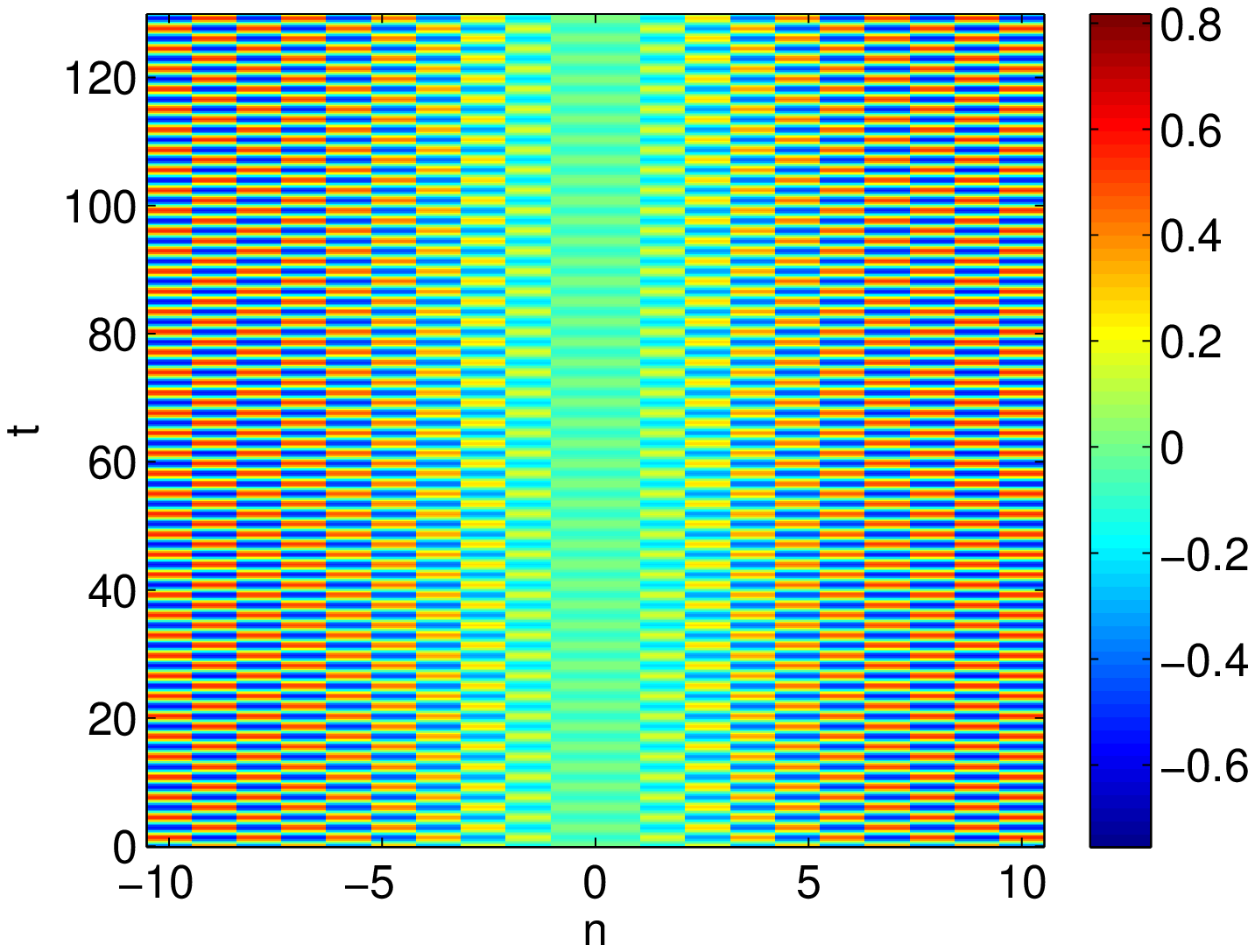,width=.45 \textwidth} 
\epsfig{file=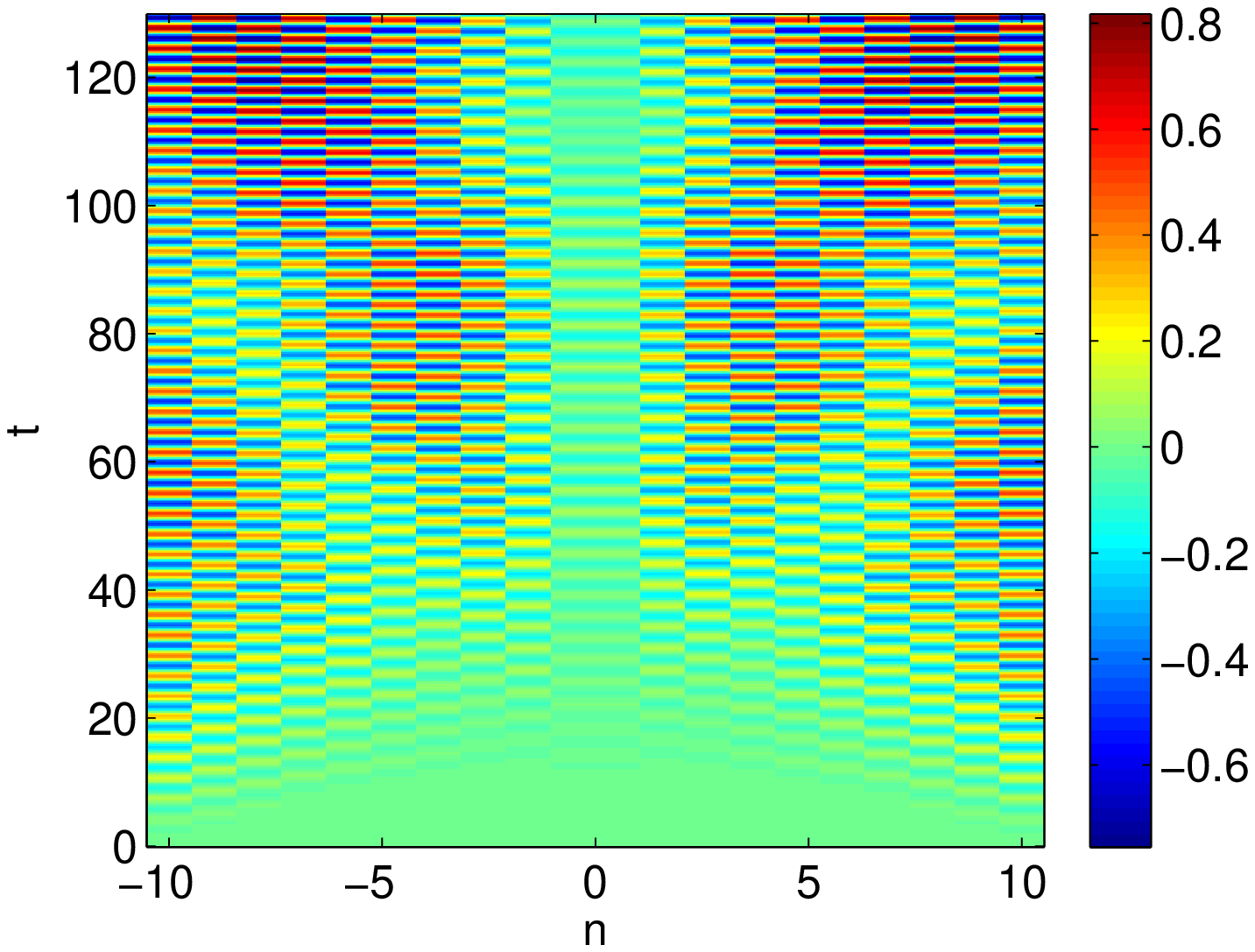,width=.45 \textwidth}}
 \centerline{\epsfig{file=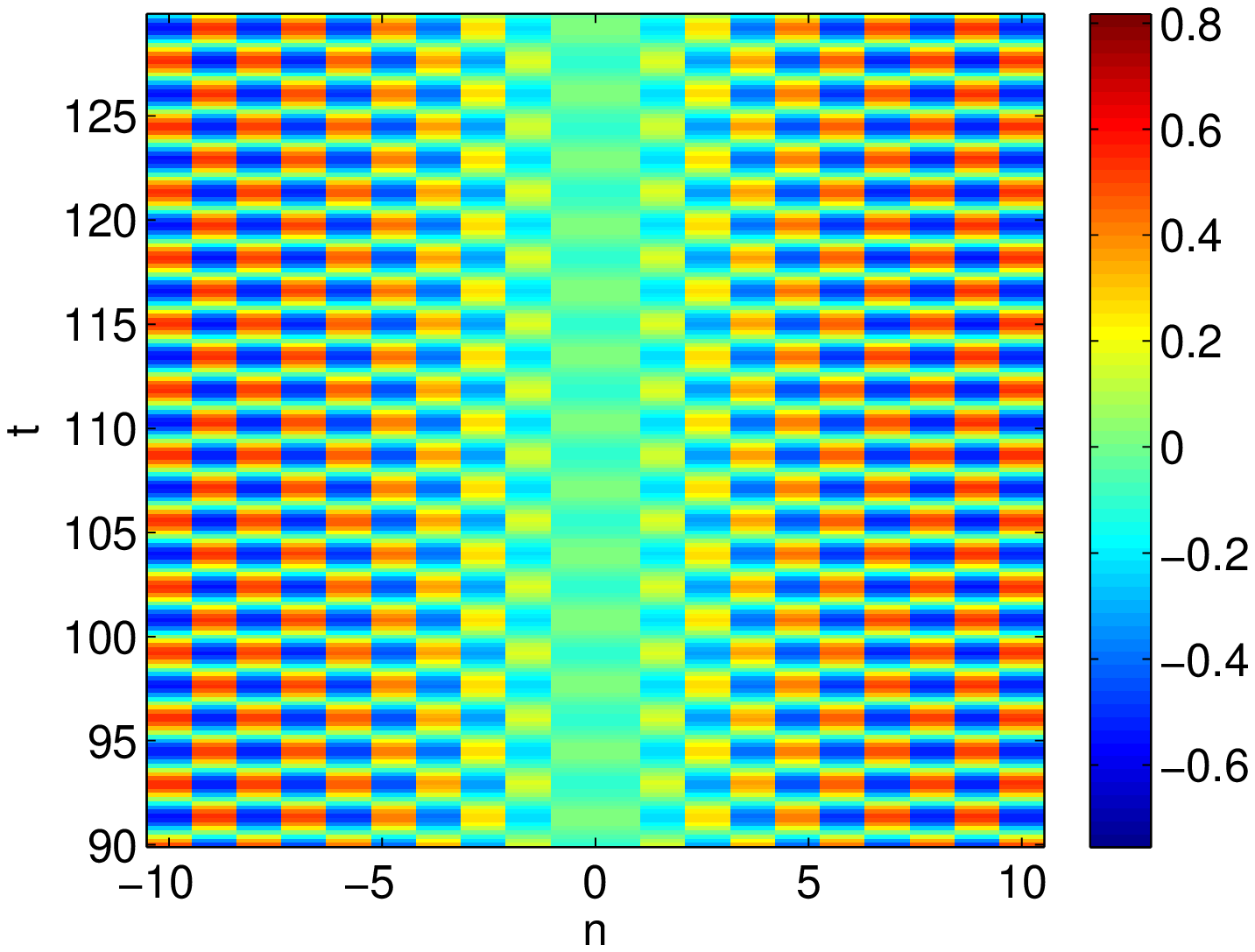,width=.45 \textwidth} 
\epsfig{file=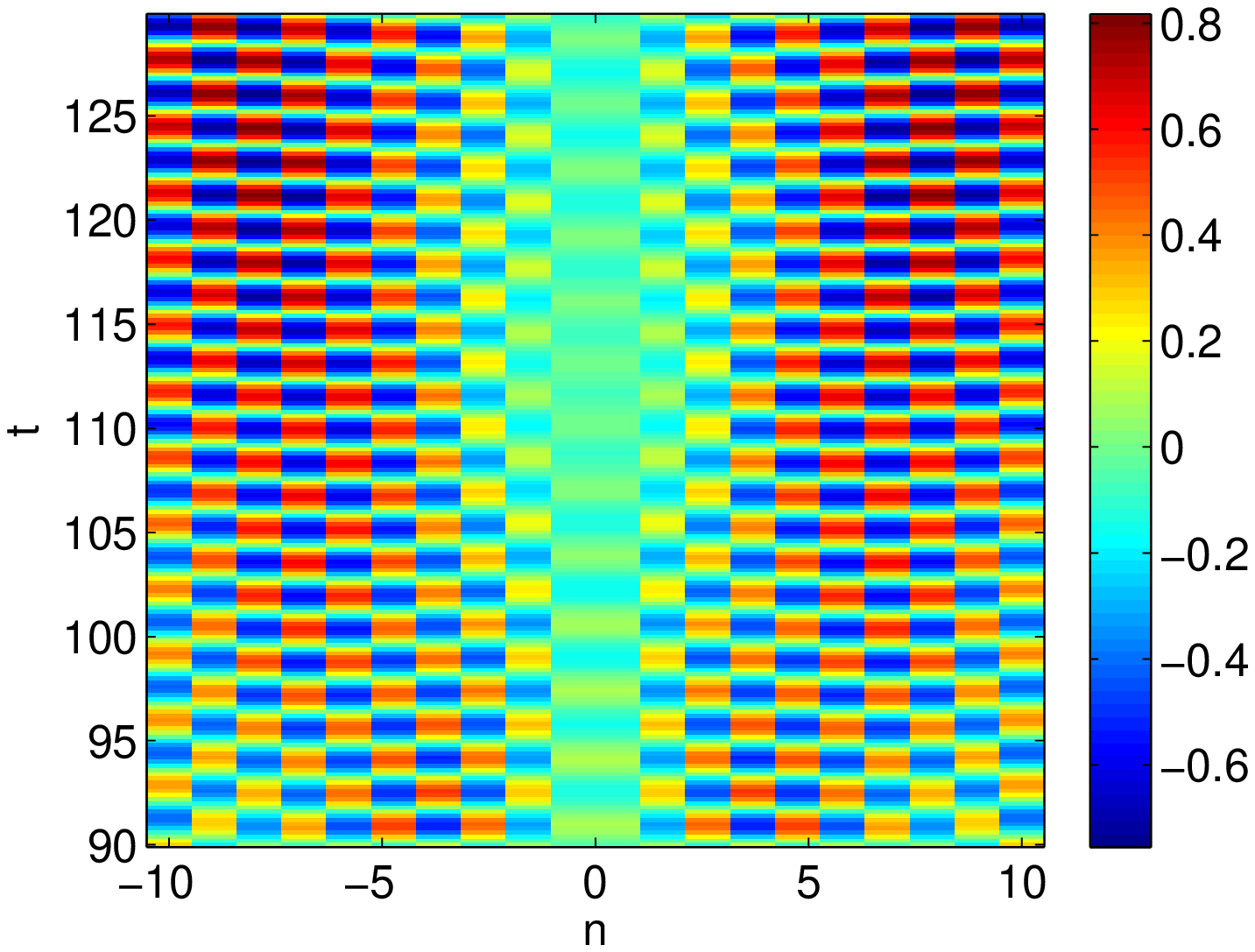,width=.45 \textwidth} }
\caption{Top: contour plots of the space-time evolution of an exact weakly nonlinear, bond-centered
dark breather with a vertical shift of $c=0$ and frequency $\omega_b=1.99$ 
(left) and an initially zero chain actuated out of phase at
the boundaries with frequency $\omega_b=1.99$ (right). The color intensity corresponds to the strain $y_n$. Bottom: A zoom of the corresponding
top panels between times $90$ and $130$. }
\label{contour}
\end{figure}

In an experimental setting, it is difficult to impose an initial displacement 
and velocity to an arbitrary number of beads,
and thus the exact solutions described above cannot be prescribed 
as an initial condition to the lattice.
Hence, we have to resort to alternative approaches taking into consideration
the somewhat limited actuation of the finite chain which is currently
available experimentally. In that light, we propose 
an  approach which relies solely on
exciting (or actuating) the ends of a finite chain,
and taking advantage of the resulting interference effect,
in order to produce a time-periodic solution with a density dip
(i.e., a dark breather).


To verify theoretically the presence of dark breathers when only the boundaries of a chain are excited, we integrate Eq.~\eqref{GC} numerically with a chain of beads at rest.
We integrate the displacement variant of the equation since this corresponds directly to the physical situation.
The numerically exact dark breathers which are in the strain variables $y_n$, can be converted by using
the relation~\eqref{strain2dis}. Since Eq.~\eqref{GC} is invariant under the transformation $u_n \rightarrow u_n + s$, with
$s \in \R$, we can vertically shift the resulting displacement dark breather by an amount $s$ 
such that the center node of the solution has zero amplitude ($u_{\mathrm{middle}}=0$). The actuators are represented by applying out of phase boundary conditions
$ u_{\mathrm{left}}(t) = a(t) \cos(\omega_b t) + s $ and $u_{\mathrm{right}} = a(t) \cos(\omega_b t + \pi) -s $, where $a(t)$ is a slowly monotonically 
increasing function that satisfies $ \sup a = \mu$, where $\mu$ is the amplitude of the exact dark
breather solution of interest with frequency $\omega_b$. By driving the chain in this way, the (resulting plane) waves propagating
from the left boundary and the right boundary will cancel out at the center 
bead resulting in a solution with $u_{\mathrm{middle}} = 0$,
as desired.
 Thus, the resulting symmetry in the displacement variables
(see bottom right panel of Fig.~\ref{profiles}) corresponds to a bond-centered solution in the strain-variables 
(see bottom  left panel of Fig.~\ref{profiles}). Since this is an interference experiment, we chose an actuating
frequency that lies within the phonon band such that linear waves will propagate through the chain. 

The left panels of Fig.~\ref{contour} show contour plots of the space-time evolution of a weakly nonlinear, bond-centered
dark breather with a vertical shift of $c=0$ in the strains. The amplitude
in this case is $\mu \approx 0.8$. The checkered pattern is a result of the time periodicity. Notice the density
dip at the center of the structure. The right panel shows the contour plot of the space-time evolution of
the actuated chain. There is a strong resemblance to a 
dark breather for time sufficiently large (here for $t>90$).

\begin{figure} 
 \centerline{\epsfig{file=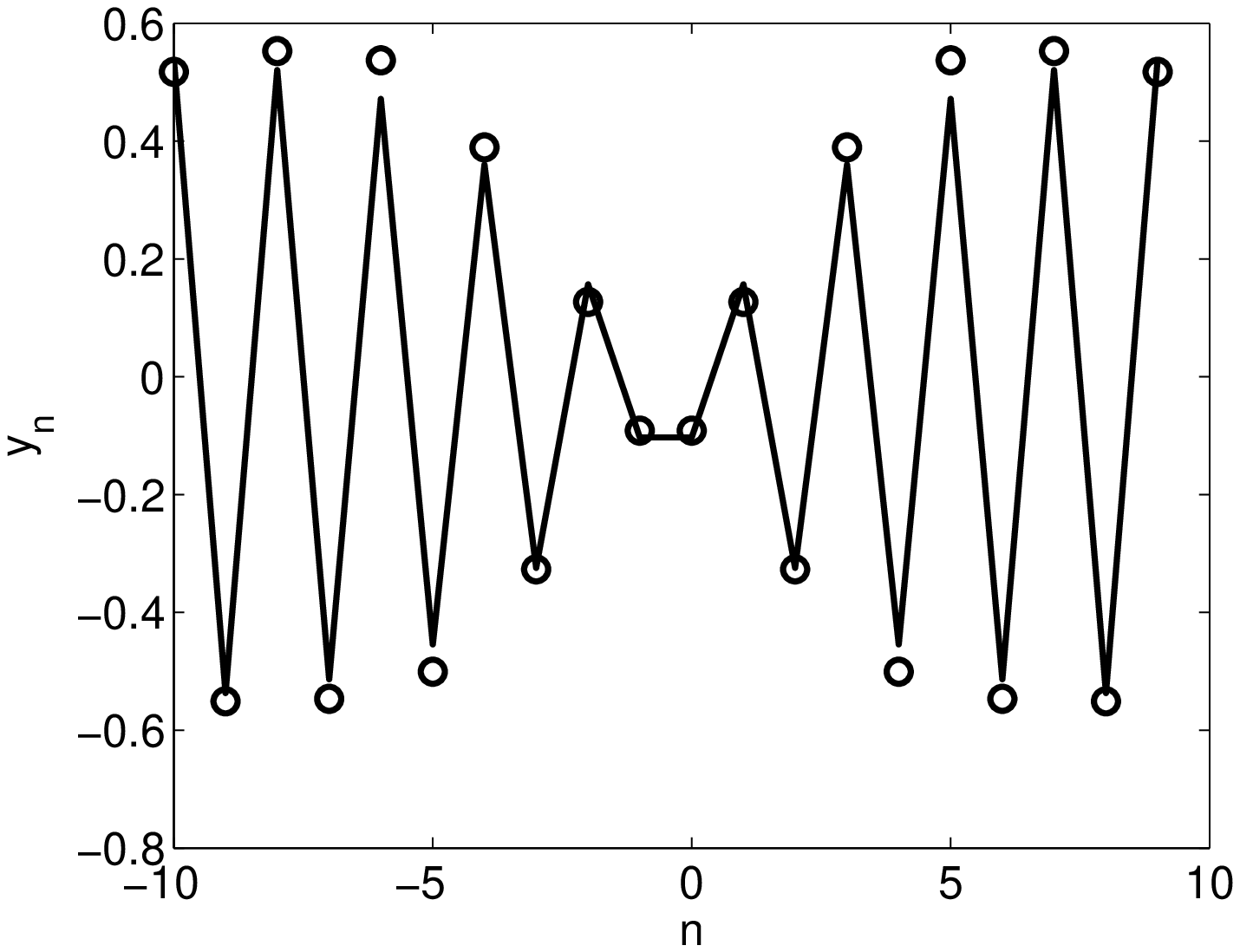,width=.45 \textwidth} 
\epsfig{file=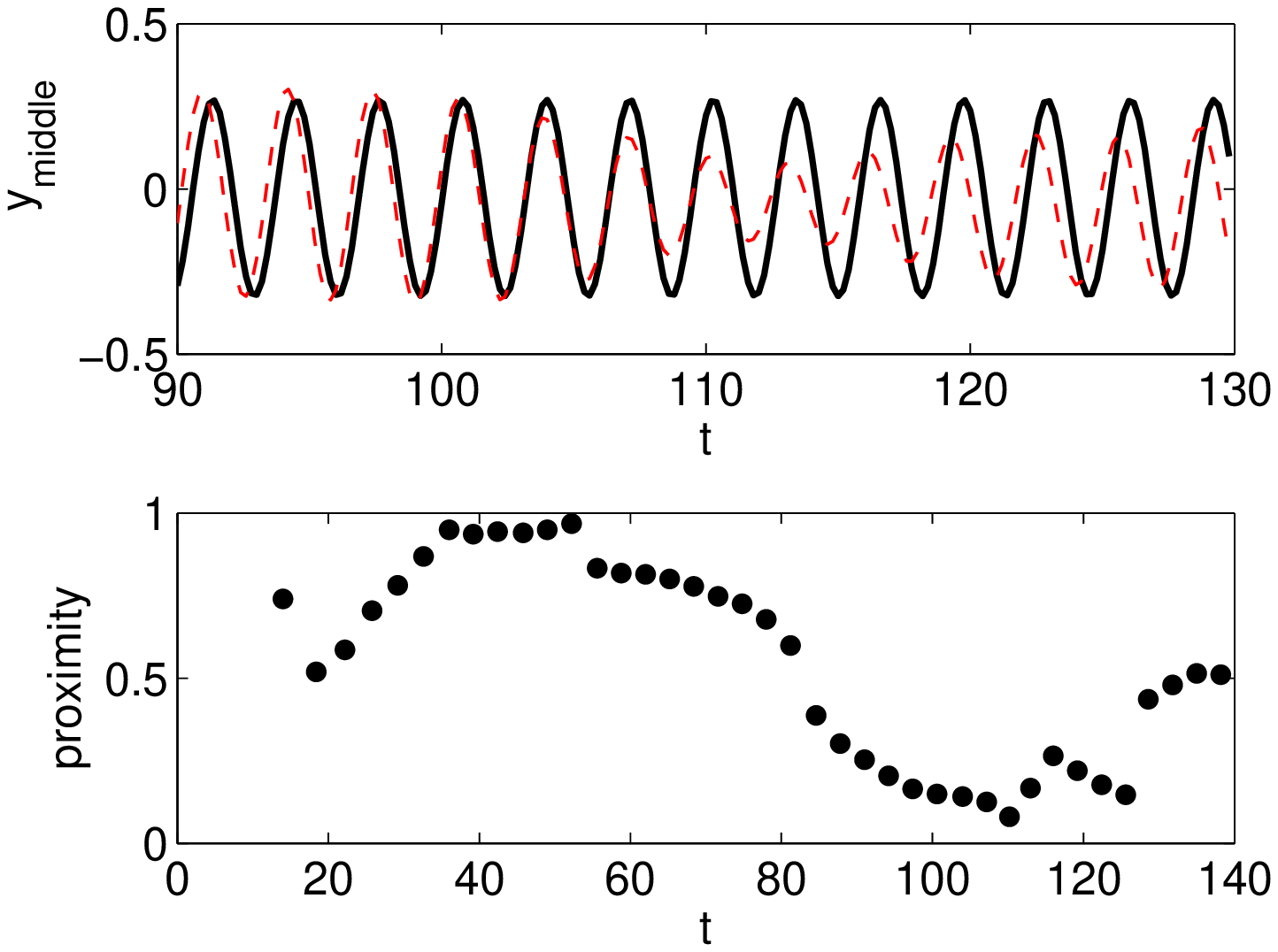,width=.45 \textwidth}}
\caption{Left: The driven chain at $t\approx 110$ (markers) and an exact dark breather (solid line) for $\omega_b=1.99$.
Top Right: Evolution of the middle bead for the driven chain (dashed line) and exact dark breather (solid line). 
Bottom Right: The $l^\infty$ norm of the difference between the driven chain and exact dark breather for time points
such that $y_{\mathrm{middle}}(t)$ is at a maximum. }
\label{proximity}
\end{figure}

To measure the proximity of the interference to an exact dark breather solution we compared the two directly.
First, we compared the difference of the time evolution of the center nodes $y_{\mathrm{middle}}(t)$ of the 
exact and approximate breathers (top right panel of Fig.~\ref{proximity}). It was also useful to
compare the entire solution at one time slice per period of oscillation 
(the slice that corresponds to a maximum of
 $y_{\mathrm{middle}}(t)$ ). We used the $l^\infty$ norm to measure difference 
of the two structures 
(bottom right panel of Fig.~\ref{proximity} ). The most ``dark breather like'' 
state during the propagation (using these measures) is shown in the left panel of Fig.~\ref{proximity}. Here, the presence of
the dark breather is clear, albeit slightly disturbed by presence of the linear waves. Essentially here, we employ linear waves to produce the destructive
interference pattern associated with the dark breathers and rely on the 
nonlinearity to subsequently produce the fundamental nonlinear state 
associated with such a setting.

\begin{figure} 
 \centerline{
\epsfig{file=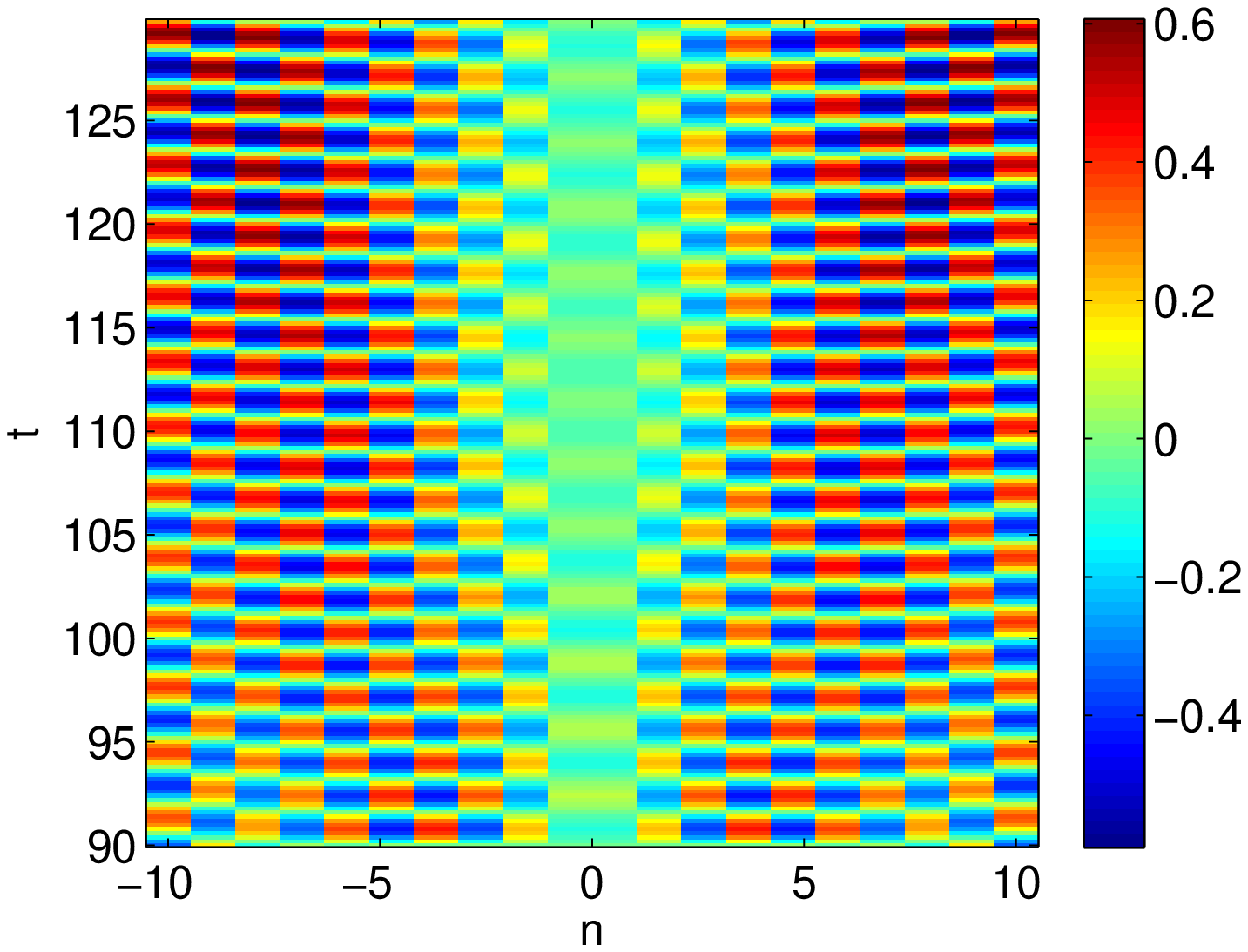,width=.45 \textwidth}
\epsfig{file=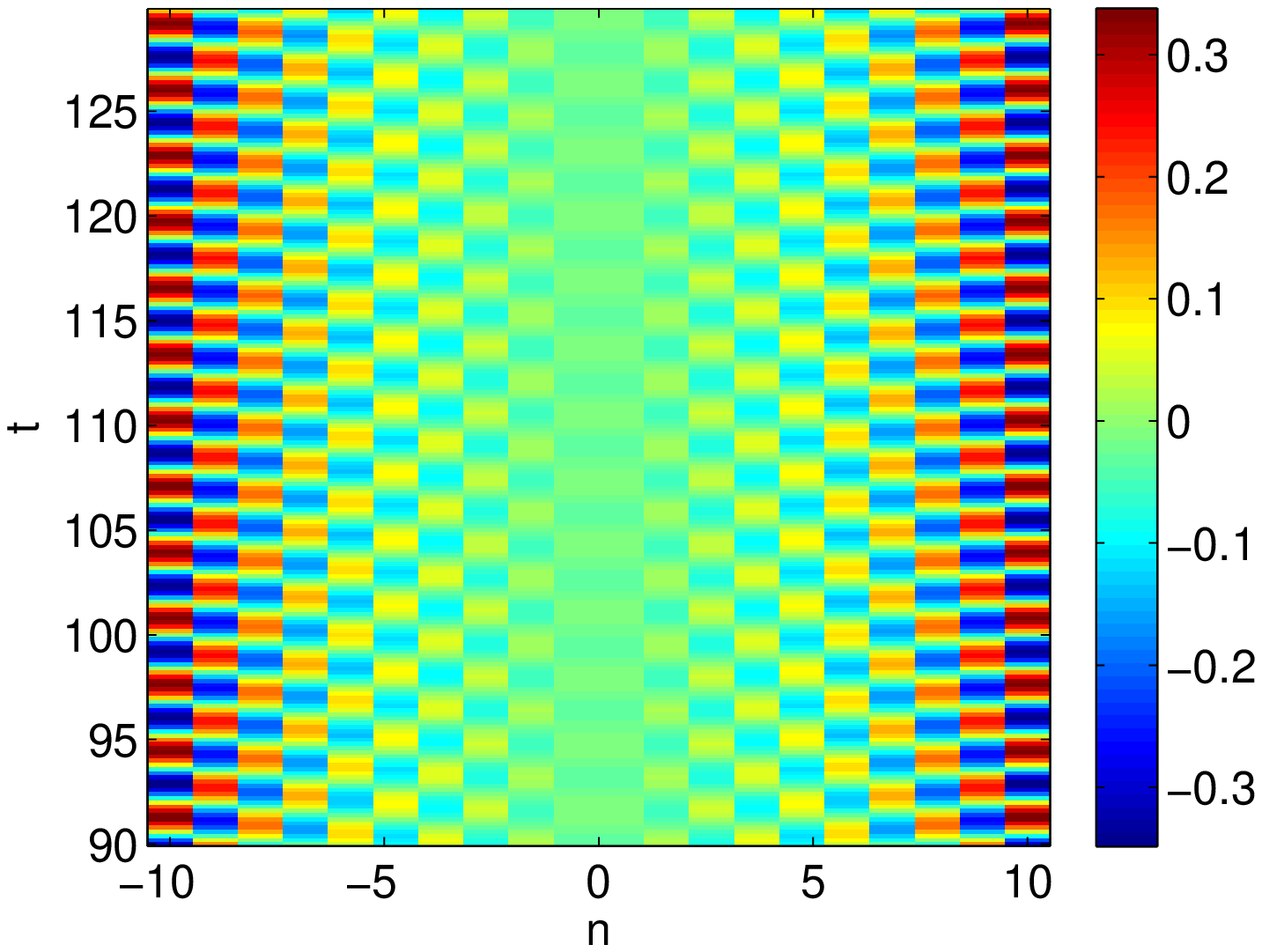,width=.45 \textwidth }}
 \centerline{
\epsfig{file=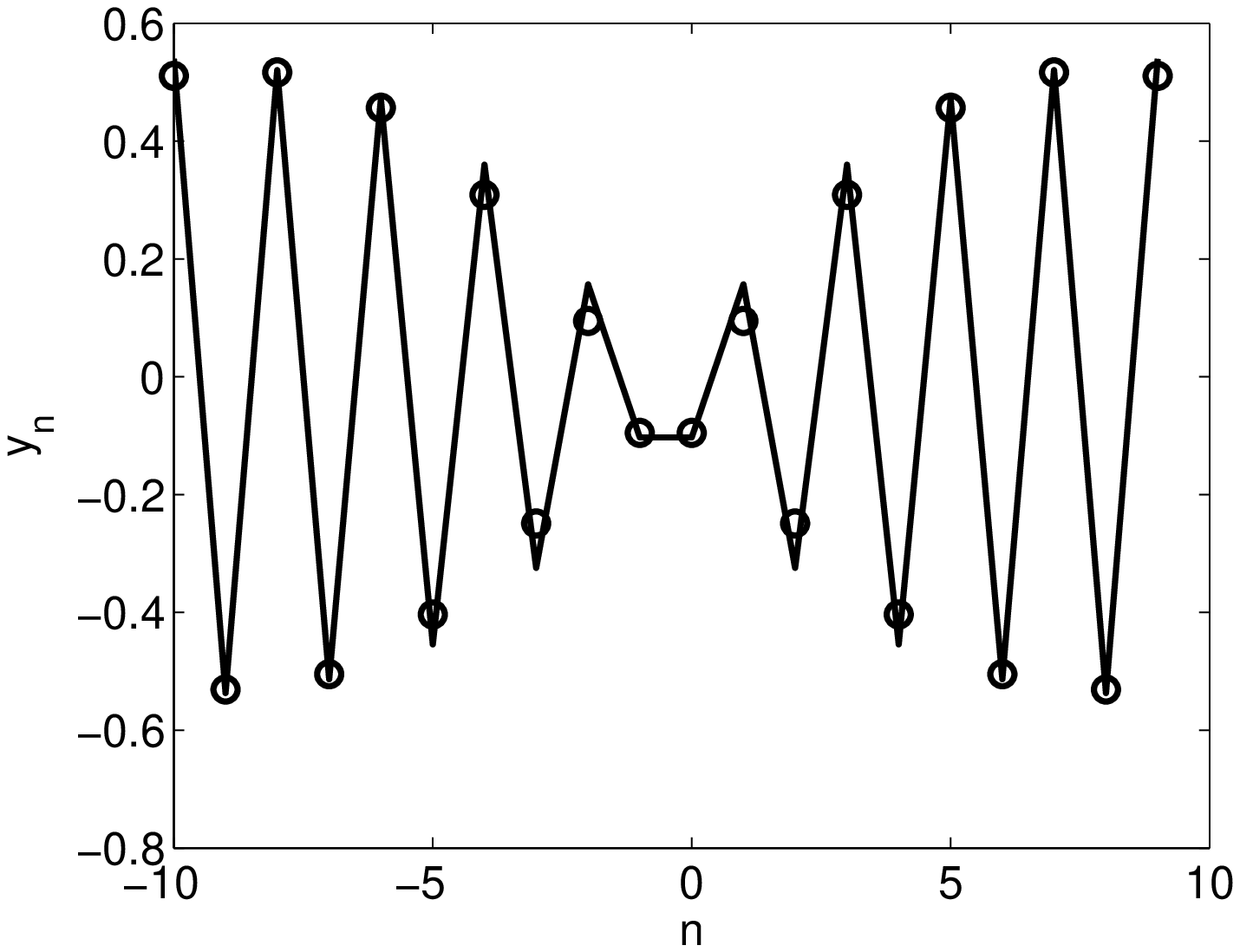,width=.45 \textwidth} 
\epsfig{file=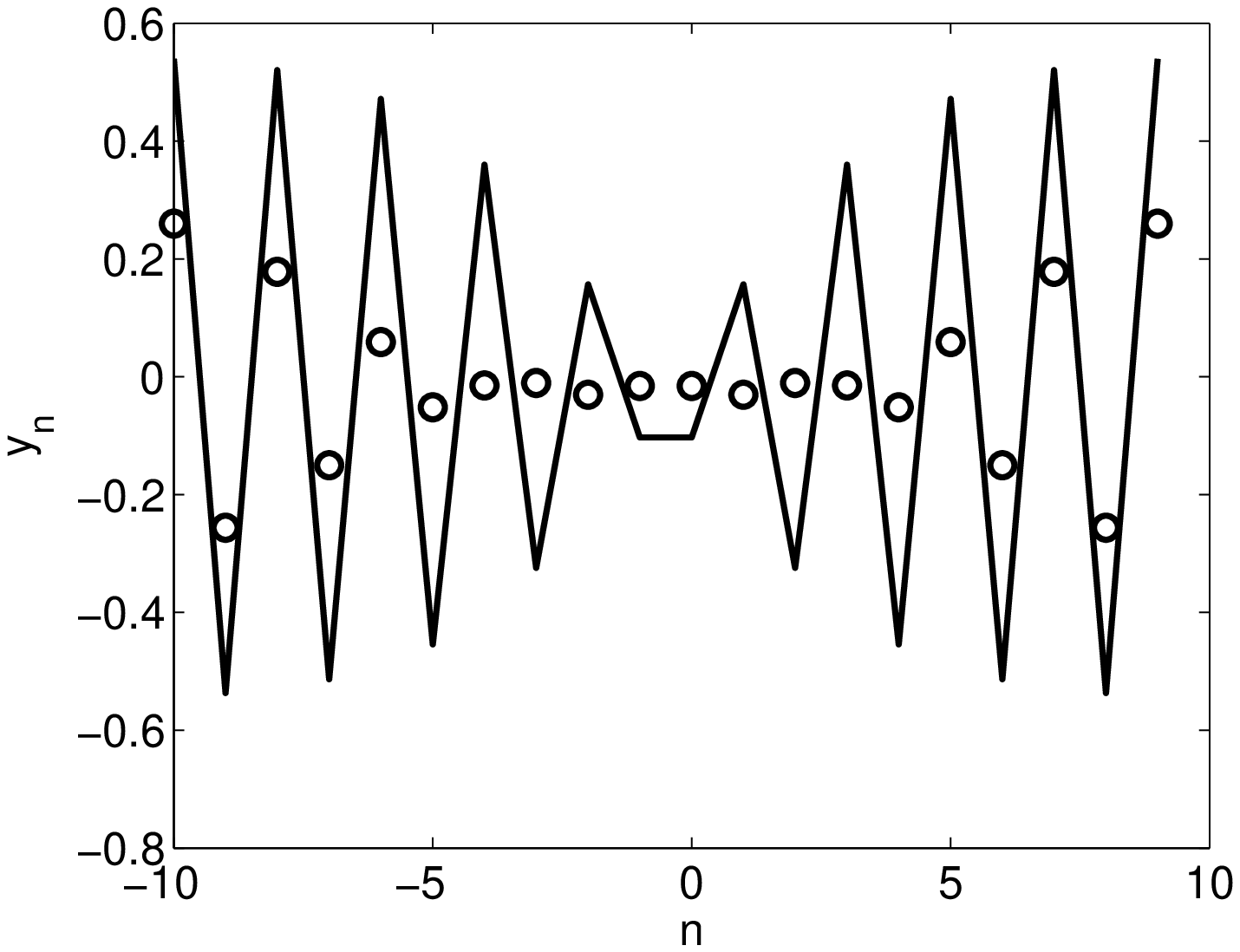,width=.45 \textwidth} } 
\caption{Top: Contour plots of the space-time evolution of the driven chain with the same parameters as in Fig.~\ref{contour}
and with a dissipation coefficient of $\tau = 100$ (left) and $\tau = 10$ (right). Bottom: Comparison between
the driven chain (markers) and exact breather solution (solid line) for $\tau = 100$ (left) and $\tau = 10$ (right). Here
the times were chosen to yield the closest resemblance. }
\label{diss_proximity}
\end{figure}

The timescale which is required for the dark breathers to manifest
via the actuation is crucial, as the physical system will have dissipative effects, which may be detrimental to the emergence of
the dark breathers. Dissipation can be modeled \cite{Nature11} by 
modifying Eq.~\eqref{GC} as
\begin{equation} \label{GC-dis}
M \ddot{u}_n = A [ \delta_0 +   u_{n-1} -   u_{n}]_+^{3/2} - A [ \delta_0 +   u_{n} - u_{n+1}]_+^{3/2}  -  \frac{M}{\tau} \dot{u}_n. 
\end{equation}
For small values of the dissipation coefficient (i.e. for large $\tau$), dark breathers can still emerge (see Fig.~\ref{diss_proximity} left), however
for larger values, the dissipative effects are indeed detrimental to the emergence of the dark breathers (see Fig.~\ref{diss_proximity} right).
Thus, it is important to consider  physical values of the parameters. The 
scaling,
 $$  u_n = \delta_0 \, \tilde u_n   , \quad \quad t =  \beta  \, \tilde t, \quad \quad \tau = \beta\, \tilde{\tau}, \quad \beta = \sqrt{ \frac{2 M}{3 A \delta_0^{1/2} }},  $$
can be used to recover the physically relevant solutions, where the variables with the tilde represent solutions in 
the idealized case of $\delta_0=M=1$ and $A=2/3$ used above. For a chain of 316-stainless steel beads we have,
$$ A = \frac{E \sqrt{2 R}}{3(1 - \nu^2)},\quad \quad M = 4 \pi R^3 \rho/3, \,  \quad \quad \delta_0 = (F_0/A)^{2/3},  $$
where $E = 193 \cdot 10^{9}$\, $(N/m^2)$  is the Young's Modulus,
$R = 9.525\cdot 10^{-3} \, (m)$ is the bead radius, $\nu = 0.3$
is the Poisson Ratio, $\rho = 8027.17$\, $(kg/m^3$) is the bead density,
and $F_0 = 20 \, (N)$ is a typical precompression force
which results in physical parameter values 
of $ A \approx  9.7576\cdot 10^{9} \, (N/m^{3/2})$,  $M \approx 0.0029 \, (kg)$  and  $\delta_0 \approx 1.6136\cdot 10^{-6} \, (m)$.
The scaling parameter in this case becomes $\beta \approx 3.9533 \cdot 10^{-5}$.
Thus, a breather appearing at approximately $\tilde{t} \approx 90 $ would correspond to  $ t = 90 \beta \approx  3.6 \,(ms)$ 
in physical time. The tested dissipation coefficients of $\tilde{\tau} = 100$ and $\tilde{\tau}=10$ would become
$\tau= 100 \beta \approx 4 \, (ms)$ and $\tau= 10 \beta \approx  0.4 \, (ms)$ respectively in the physical scaling.
Thus, if one were able to carry out experiments where the dissipation constant is on the order of 
$\tau= 4 \, (ms)$, dark breathers should be detectable. We note that the experimentally relevant dissipation constant 
in Ref.~\cite{Nature11} corresponds to $\tau = 2 \, (ms)$,  which demonstrates the
feasibility of detecting dark breathers in currently
available experimental settings of granular chains.

\section{Discussion and future directions}

We have demonstrated that dark breathers are not only an interesting entity
of the granular crystal model from a theoretical point of view, but are also 
within the reach of currently available experimental technology, given
the ability to actuate both ends of the chain of beads. On the theoretical
side, a useful qualitative (although only approximate) approach towards
the understanding of such states arises through the use of the NLS 
reduction. On the other hand, their detailed numerical study illustrates
the existence of a small amplitude branch arising from one of the linear
eigenmodes of the system, but also of a large amplitude branch 
which cannot be captured from our analytical considerations. The destructive
interference of the signals (at the appropriate frequency inside the pass
band) induced by the two actuators produces a waveform proximal to the
exact dark breathers, and this observation is found to persist even in
the presence of realistic values of the dissipation within the system.

 Besides the realization of
such experiments which would be of particular value, our study also raises 
several interesting open problems from the mathematical/theoretical point
of view. Among these, we highlight 
a band analysis of the Floquet multipliers in order to capture the
continuous spectrum of the infinite lattice problem; also, 
a proof of the numerically observed nonlinear instability, 
apparently present due to a non-trivial
scale invariance that we identified; finally, the rigorous
justification of the NLS equation for dark breathers
and especially of its regime of validity (which in our case appears
to be relatively limited at least in a quantitative sense).  

The readiness of the solutions to move, via random perturbations of the 
unstable solutions, and the apparent 
spatial translation invariance of the system render also plausible 
the potential existence of traveling breathers. In that light, 
a computational study of such states along the lines of earlier similar
studies in the realm of the discrete NLS equation~\cite{floria,our1,hadi}
would be of particular interest.
The potential emergence of dark breathers  
in inhomogeneous chains, such as dimers~\cite{Boechler2009,Theo2010} or 
trimers~\cite{apl11} may be another issue to computationally
consider in detail, since bright
breathers have been identified as relatively robust nonlinear
states in such settings~\cite{Boechler2009,Theo2010,hooge11}.

Such studies are currently in progress and will be reported in 
future publications.

\section*{Acknowledgements}
The authors would like to thank Guillaume James, Jes\'us Cuevas, Bernardo S\'anchez Rey,
Roy Goodman, and Mason A. Porter for useful discussions.
CD acknowledges support from the National Science Foundation, grant 
number CMMI-844540 (CAREER) and NSF 1138702. PGK gratefully acknowledges
support from the US National Science Foundation under
grant CMMI-1000337, the US Air Force under grant 
FA9550-12-1-0332, the Alexander von Humboldt Foundation, as well
as the Alexander S. Onassis Public Benefit Foundation.

\end{document}